\documentstyle [12pt,epsf] {article}

\begin{document}

\begin{titlepage}

\begin{center}
\vspace{2cm}
\LARGE
The Host Galaxies of AGN                            
\\                                                     
\vspace{1cm} 
\large
Guinevere Kauffmann$^1$, Timothy M. Heckman$^2$, Christy Tremonti$^2$,\\ 
Jarle Brinchmann$^1$,   
St\'ephane Charlot$^{1,3}$, Simon D.M. White$^1$ ,\\ 
Susan E. Ridgway$^2$,
Jon Brinkmann$^4$, Masataka Fukugita$^5$,\\ 
Patrick B. Hall$^{6,7}$, \v{Z}eljko Ivezi\'{c}$^7$,
Gordon T. Richards$^7$, Donald P. Schneider$^8$\\
\end{center}

\vspace{0.3cm}
\small
\noindent
{\em $^1$Max-Planck Institut f\"{u}r Astrophysik, D-85748 Garching, Germany} \\
{\em $^2$Department of Physics and Astronomy, Johns Hopkins University, Baltimore, MD 21218}\\
{\em $^3$ Institut d'Astrophysique du CNRS, 98 bis Boulevard Arago, F-75014 Paris, France} \\
{\em $^4$ Apache Point Observatory, P.O. Box 59, Sunspot, NM 88349} \\
{\em $^{5}$ Institute for Cosmic Ray Research, University of Tokyo, Chiba
277-8582, Japan}\\
{\em $^{6}$ Departamento de Astronom\'{\i}a y Astrof\'{\i}sica, Facultad
de F\'{\i}sica, Pontificia Universidad Cat\'{o}lica de Chile,
Casilla 306, Santiago 22, Chile}\\
{\em $^{7}$ Princeton University Observatory, Peyton Hall, Princeton NJ 08544-1001}\\
{\em $^{8}$ Department of Astronomy and Astrophysics, 525 Davey Laboratory, Pennsylvania
State University, University Park, PA 16802}\\

\vspace{0.6cm}
\large       
\begin {abstract}
We examine the properties of the host galaxies of  22,623 narrow-line  AGN with $0.02<z<0.3$ selected from a 
complete sample of 122,808 galaxies from  the Sloan Digital Sky Survey. 
We focus on the luminosity of the [OIII]$\lambda$5007  emission line as a tracer 
of the strength of activity in the nucleus.  We  study how AGN host properties
compare to those of normal galaxies and how they depend on L[OIII].
We find that AGN of all luminosities  reside almost exclusively
in massive galaxies and have distributions of  sizes, stellar surface
mass densities and concentrations that are similar to those of  ordinary early-type 
galaxies in our sample. The host galaxies of low-luminosity AGN have stellar populations 
similar to normal early-types.  The hosts of high-luminosity AGN have much younger
mean stellar ages. The young stars are not preferentially located
near the nucleus of the galaxy, but are spread out over scales of
at least several kiloparsecs. 
A significant fraction of high-luminosity
AGN have strong H$\delta$ absorption-line equivalent widths, indicating
that they experienced a burst of star formation in the recent past.  
We have also examined the stellar populations of the host galaxies
of a sample of broad-line  AGN. We conclude that there is
no significant difference in stellar content between type 2 Seyfert
hosts and QSOs with the same [OIII] luminosity and redshift. 
This establishes that a young stellar population is a general property of 
AGN  with
high [OIII] luminosities.

\end {abstract}
\vspace {0.8 cm}
\end {titlepage}
\normalsize

\section{Introduction}

The discovery of a tight correlation between black hole mass and bulge velocity dispersion
(Ferrarese \& Merritt 2000; Gebhardt et al 2000) gives  credence to theoretical arguments that spheroid 
formation and the growth of supermassive black holes are closely linked
(e.g. Richstone et al 1998; Haehnelt, Natarajan \& Rees 1998; Kauffmann \& Haehnelt 2000;
Monaco, Salucci \& Danese 2000). Active galactic nuclei (AGN) are believed to be powered by 
the accretion of gas onto black holes located  at the centres of galaxies (Lynden-Bell 1969).
AGN thus signpost galaxies in which black holes are  forming and it is
natural to hypothesize  that at least some AGN should be found in young 
spheroids. Direct observational evidence in support of this hypothesis
remains  controversial, however.

Recent  Hubble Space Telescope imaging studies of QSO host galaxies at
low redshifts have found that the majority of the most  luminous QSOs
reside in early-type galaxies (e.g.
McLure et al (1999); Bahcall et al (1997) find that around 15\% of the
quasars in their sample are found in spirals).  Although a subset of quasar host galaxies
exhibit disturbed morphologies (Percival et al 2001), the stellar populations
of these hosts appear indistinguishable from those
of normal elliptical galaxies (McLure, Dunlop \& Kukula 2000; Nolan et al 2001).

The claim that luminous QSOs are located in old ellipticals sits uncomfortably
with the results of several major spectroscopic investigations that
have confirmed the presence of young stars in at least half of  all powerful 
narrow-line or type 2 Seyfert nuclei
(Schmitt, Storchi-Bergman \& Cid Fernandes 1999; Gonzalez Delgado, Heckman \& Leitherer 2001;
Cid Fernandes et al. 2001; Joguet et al 2001). 
There is also  considerable evidence that the properties of AGN  
hosts  depend strongly on the luminosity of the central source.
The lowest luminosity AGN, the  LINERs, have 
[OIII]$\lambda$5007 and H$\alpha$ narrow-line region (NLR)  emission line luminosities
in the range $\sim 10^5$ to $10^6 L_{\odot}$. LINERs are found in galaxies of earlier Hubble
type than Seyferts and their nuclear continua are usually dominated                            
by old stars (Heckman 1980a,b; Ho, Filippenko \& Sargent 2003).

Up to now, studies  of AGN host galaxies have been limited
by small sample size. In order to carry out detailed  statistical analyses of host galaxy properties,
one requires complete  magnitude-limited samples of galaxies 
with spectra of high enough quality to identify AGN based on the       
characteristics of their emission lines. A catalog of 26 Seyfert 1, 23 Seyfert 2 and 33 LINERs
identified in the CfA Redshift survey (Huchra \& Burg 1992) has formed the basis of many follow-up
studies. De Grijp et al (1992) identified  $\sim 220$ Seyfert galaxies  from a sample of
563 IRAS sources selected from the Point Source Catalog.
Most recently, Ho, Filippenko \& Sargent (1995)  carried out a survey of
486 nearby bright  galaxies from the Revised Shapley-Ames Catalog with $B < 12.5$.
The proximity of the galaxies in this survey meant that very high spatial resolution
of the nuclear regions of the galaxies could be achieved. In addition, the spectra were of
high quality, enabling the detection of relatively weak emission lines.
Ho et al. found that  43\% of the objects in their sample could be  classified as
AGN, a significantly higher fraction than in previous surveys of galaxies selected at optical
wavelengths.

This paper examines the properties of 22,623 narrow-line AGN  selected from 
a complete sample of 122,808 galaxies from  the
Sloan Digital Sky Survey with r-band magnitudes in the range $14.5 < r < 17.7$. 
The relations between stellar mass, star formation history, size and internal structure
for the galaxies in the parent sample are described in a recent paper by   Kauffmann et al (2003b, hereafter
Paper II) where it was found
that ordinary galaxies divide into two distinct families at a stellar mass of $3 \times 10^{10} M_{\odot}$.
Low mass galaxies have young stellar populations, low surface mass densities, and the
low concentrations typical of disks. At stellar masses above $3 \times 10^{10} M_{\odot}$, 
a rapidly increasing fraction of galaxies has old stellar populations, high surface
mass densities and the high concentrations typical of bulges. The star formation histories
of galaxies correlate most strongly with surface mass density, with a 
transition from ``young'' to ``old'' stellar populations occurring 
at $\mu_* \sim 3 \times 10^8$ $M_{\odot}$
kpc$^{-2}$. 

In this paper, we compare  the properties of AGN hosts with those
of normal galaxies. The properties that we study include stellar masses,
sizes, and surface densities, as well as stellar ages and past star formation histories,
as deduced from key spectral features such as 4000 \AA\
break strengths and the equivalent widths of Balmer
absorption lines. We also study how these properties vary as a function of AGN luminosity
as measured by  the strength of the [OIII]$\lambda$5007 emission line. Section 2
reviews the properties of the galaxy sample and the methods used to derive parameters such
as stellar mass, dust attenuation strength and burst mass fraction. In section 3, we describe
how type 2 AGN are identified and classified and we investigate whether ``featureless'' nonstellar continuum
can affect our conclusions about  stellar ages and star formation histories. In section 4, 
we present a step-by-step comparison of the properties of type 2  AGN hosts with those of normal galaxies.
Section 5, compares the stellar populations of
the most powerful type 2 AGN in our sample with those of type 1 AGN
of the same [OIII] luminosity and redshift.
Finally, in section 6, we show images of representative examples of our most powerful AGN
and  discuss the implications of our results.

Throughout this paper we assume a Friedman-Robertson-Walker cosmology with
$\Omega=0.3$, $\Lambda=0.7$ and $H_0$=70 km s$^{-1}$ Mpc$^{-1}$.

\section {Review of the Spectroscopic Sample of Galaxies}

The Sloan Digital Sky Survey
(York et al. 2000; Stoughton et al. 2002)
is using a  dedicated 2.5-meter wide-field
telescope at the Apache Point Observatory to conduct an imaging and
spectroscopic survey of about a quarter of the extragalactic sky. The imaging is
conducted in the $u$, $g$, $r$, $i$, and $z$ bands (Fukugita et al. 1996; Gunn et al. 1998;
Hogg et al. 2001; Smith et. al. 2002),
and spectra are obtained with a pair of multi-fiber spectrographs.
When the current survey is complete, spectra will have
been obtained for nearly 600,000 galaxies and 100,000  QSOs selected uniformly
from the imaging data. Details on the spectroscopic target selection
for the ``main'' galaxy sample and QSO sample can be found in
Strauss et al. (2002) and
Richards et al. (2002) respectively. Details about the tiling algorithm
and the astrometry can be found in Blanton et al (2003) and Pier et
al (2003), respectively.  The results in this paper are   
based on spectra of $\sim$122,000 galaxies with 
$14.5 < r < 17.77$  contained in the the SDSS Data
Release One (DR1). These data are
to be made publicly available in 2003.

The spectra are obtained
through 3 arcsec diameter fibers. At the median redshift of the main
galaxy sample ($z \sim$ 0.1), the projected aperture diameter  is
5.5 kpc and  typically contains 20 to 40\% of
the total galaxy light. The SDSS spectra are thus closer to global
than to nuclear spectra. At the median redshift the spectra
cover the rest-frame wavelength range from $\sim$3500 to 8500 \AA\
with a spectral resolution $R \sim$ 2000 ($\sigma$$_{instr} \sim$
65 km/s). They are spectrophotometrically calibrated through
observations of F stars in each 3-degree field.
By design, the spectra are well-suited to the determination
of the principal properties of the stars and ionized gas in galaxies.
The absorption line indicators (primarily the 4000 \AA \hspace{0.1cm} break strength and
the H$\delta_A$ index) and the emission line fluxes analyzed in this paper are calculated using a 
special-purpose code described in detail in Tremonti et al (2003, in preparation).
A detailed description of the galaxy sample and the methodology used to  derive
parameters such as stellar mass and dust attenuation strength
can  be found in Kauffmann et al (2003a, hereafter Paper I).

The rich stellar absorption-line spectrum of a typical SDSS galaxy provides 
unique information about its stellar content and  dynamics. 
However, it makes the measurement
of weak nebular emission-lines quite difficult. To deal with this,
we have performed a careful subtraction of the stellar absorption-line
spectrum before measuring the nebular emission-lines. This is accomplished
by fitting the emission-line-free regions of the spectrum with a model
galaxy spectrum computed using   the new population synthesis
code of Bruzual \& Charlot (2003, hereafter BC2003), which incorporates
a high resolution (3 \AA\ FWHM) stellar library. A set of 39 model template
spectra were used  spanning a wide range  in age and metallicity.
After convolving the template spectra to the measured stellar
velocity dispersion of an
individual SDSS galaxy,
the best fit to the galaxy spectrum is constructed from
a non-negative linear combination of the template spectra.

We have used the amplitude of the 4000 \AA\ break (the narrow version of the index defined in
Balogh et al. 1999)
and the strength of the H$\delta$ absorption line (the Lick
$H\delta_A$ index of Worthey \& Ottaviani 1997) as diagnostics of the
stellar populations of the host galaxies. Both indices
are corrected for the observed contributions of the emission-lines
in their bandpasses.
Using a library of
32,000 model star formation histories,
we have used the measured $D_n(4000)$ and $H\delta_A$ indices
to obtain a maximum likelihood  estimate of  the  $z$-band mass-to-light ratio for each galaxy.
By comparing the colour predicted by the  best-fit model to the observed colour of the galaxy,
we also estimate the attenuation of the starlight due to dust.

The SDSS imaging data provide the basic structural parameters that are used in this analysis.
We use the $z$-band as our fiducial filter because it is the least sensitive to the
effects of dust attenutaion.
The $z$-band absolute magnitude, combined with our estimated values of M/L and dust attenuation
$A_z$ yield the stellar mass ($M_*$). The half-light radius in the $z$-band and the
stellar mass yield the effective stellar surface mass-density
($\mu_* = M_*/2\pi r_{50,z}^2$). As a proxy for Hubble type we use
the SDSS ``concentration'' parameter $C$, which is defined as the ratio
of the radii enclosing 90\% and 50\% of the galaxy light in the $r$ band
(see Stoughton et al. 2002). Strateva et al. (2001) find that galaxies
with $C >$ 2.6 are mostly early-type galaxies, whereas spirals and irregulars
have 2.0 $< C <$ 2.6.

\section{Identification and Classification of AGN }

According to the standard ``unified'' model
(e.g. Antonucci 1993), AGN can be broadly classified into two categories
depending on whether the central black hole and its associated continuum
and broad emission-line region is
viewed directly (a ``type 1'' AGN) or is obscured by a dusty circumnuclear
medium (a ``type 2'' AGN). Since this obscuring medium does not fully cover
the central source, some of the radiation escapes and photoionizes surrounding
gas, leading to strong narrow permitted and forbidden emission lines from
the `Narrow Line Region' (NLR). In type 1 AGN the optical continuum is dominated
by non-thermal emission, making it  a challenge to study the host galaxy and its
stellar population. This is especially true of QSOs, where the continuum
radiation from the central source outshines the stellar light from the host
galaxy.

We have therefore excluded the type 1 AGN from our initial sample
(but will undertake a limited analysis of such objects in section 5).
The rejection of type 1 AGN from our sample is accomplished automatically by 
the SDSS spectral classification algorithm, which is based on
a Principal Component Analysis (PCA) approach (Schlegel et al, in preparation).
We have verified the reliability of this
procedure through the manual inspection of $\sim$1000 spectra of the
most powerful AGN in our sample. In about 8\% of the cases, weak broad
wings are present on the H$\alpha$ emission-line, but not on H$\beta$.
In the standard nomenclature (Osterbrock 1989)
these would be classified as ``type 1.9'' AGN (objects in which the obscuration
of the central continuum source is substantial, but not complete).
We retain these
objects in our sample, since the contribution to the observed continuum by
the AGN is not significant (see section 3.2 below).

Baldwin, Phillips \& Terlevich (1981, hereafter BPT) demonstrated that it was possible to distinguish
type 2 AGNs from normal star-forming galaxies by considering the intensity ratios of two pairs
of relatively strong emission lines, and this technique was refined by Veilleux \& Osterbrock
(1987). It has become standard practice to classify objects according to
their position on the so-called BPT diagrams. Fig. 1 shows an example of such a diagram
for all the emission-line galaxies in our sample. We have plotted the ratio
[OIII]$\lambda$5007/H$\beta$ versus the ratio [NII]$\lambda$6583/H$\alpha$ for all galaxies
where all
four lines were detected with $S/N>3$. Note that these ratios are almost completely
insensitive to reddening or to errors in the spectrophotometry. 
This sample includes 55,757 objects (45.4\% of
our total sample of 122,808 galaxies). The other BPT diagrams involving the ratios
[SII]$\lambda\lambda$6717,6731/H$\alpha$ and [OI]$\lambda$6300/H$\alpha$ 
include a somewhat smaller fraction of objects
at a similar cut in signal-to-noise.

The exact demarcation between star-forming galaxies and AGN is subject to considerable
uncertainty. Recently, Kewley et al (2001) used a combination of 
photo-ionization and stellar population
synthesis models to place a theoretical upper limit on the location of star-forming models in the
BPT diagrams. Their models allow for a wide range in metallicity, ionization parameter
and dust depletion and also make allowances for the effects of shock excitation by supernovae.
The Kewley et al demarcation between starbursts and AGN (shown as a dotted line in Fig. 1) represents 
a very conservative {\em lower} limit on the true number of AGN in our sample.
Galaxies with emission line ratios that place them above this line cannot be explained by any possible
combination of parameters in a star-forming model. 

In practice, star-forming galaxies exhibit strong correlations between properties
such as ionization parameter and metallicity, which cause   
them to exhibit rather little scatter around a single relation in the  BPT diagram. 
This is evident from  Fig. 1, which shows that there are two well-separated sequences of
emission line galaxies and that the AGN sequence separates from the
sequence of star-forming galaxies well below the Kewley et al demarcation curve.
Based on these data, we have chosen to revise the demarcation  between starburst
galaxies and AGN  as follows: A galaxy is defined to be an AGN if
\begin{equation} \log ([\rm{OIII}]/H\beta) >  0.61/(\log ([\rm{NII}]/H\alpha)-0.05) +1.3 \end{equation}    
This curve is represented by a dashed line in Fig.1 and in what follows, we use this as our
canonical division between star-forming galaxies and AGN in our sample.

\begin{figure}
\centerline{
\epsfxsize=13cm \epsfbox{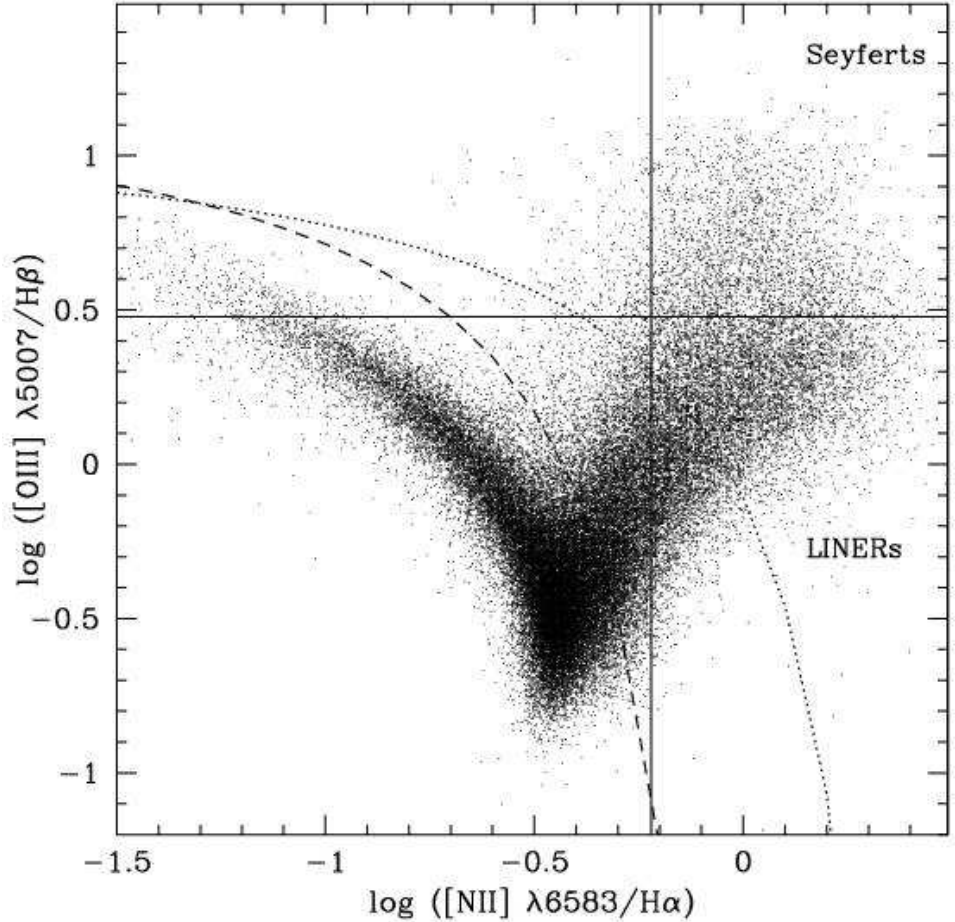}
}
\caption{\label{fig1}
\small
An example of a BPT (Baldwin, Phillips \& Terlevich 1981) diagram in which we plot
the emission line flux ratio [OIII]/H$\beta$ versus the ratio [NII]/H$\alpha$ for all
the galaxies in our sample where all four lines are detected with $S/N >3$ (55,757 objects).
The dotted curve shows the demarcation between starburst galaxies and AGN
defined by Kewley et al (2001). The dashed curve shows our revised demarcation (equation 1). 
A total of 22,623 galaxies lie above dashed curve.
Seyfert galaxies are often defined to have [OIII]/H$\beta > 3$ and 
[NII]/H$\alpha>0.6$, and LINERs to have [OIII]/H$\beta < 3$ and [NII]/H$\alpha > 0.6$.
Our sample includes 2537 Seyferts and 10,489 LINERs according to this definition.}
\end {figure}
\normalsize

Narrow-line (type 2)
AGN are traditionally divided into  3 general classes:
type 2 Seyferts, LINERs, and the so-called ``transition'' objects. 
LINERs (Heckman 1980) typically have
much lower nuclear luminosities than Seyferts. Spectroscopically, they resemble Seyferts except
that low ionization lines such as [OI]$\lambda$6300 and [NII]$\lambda \lambda$6548,6583 are
relatively strong. The relationship of LINERs to the more powerful Seyfert galaxies and QSOs
has been a topic of substantial controversy. Like the Seyferts, the emission line properties
of LINERs can be explained by ionization by a ``hard'' power-law spectrum. LINER-type spectra can, 
however, also be produced in cooling flows, starburst-driven winds and shock-heated gas
(Heckman 1987; Filippenko 1992). This has led to some debate as to whether LINERs should
be considered as a true low-luminosity extension of the AGN sequence.
Finally, Ho et al (1993) introduced a class of ``transition'' objects with nuclear emission-line 
properties intermediate between those of normal star-forming galaxies and those of Seyferts
and LINERs. Ho et al proposed that these objects are in fact ordinary LINER/Seyfert
galaxies whose integrated spectra are diluted or contaminated by neighbouring HII regions.

In traditional classification schemes (see for example Ho, Filippenko \& Sargent 1997), Seyfert galaxies
are identified as those objects with high values of both [OIII]/H$\beta$ ($>3$) and of other ratios 
involving lower ionization lines, such as [NII]/H$\alpha$, [SII]/H$\alpha$ and [OI]/H$\alpha$.
LINERs, on the other hand, have lower values of [OIII]/H$\beta$ ($< 3$), but high values of
ratios involving the lower ionization lines.

\subsection {AGN as a sequence in [OIII] Line Luminosity}

A contribution to the emission-line spectrum by
both star-formation and an AGN is almost inevitable in many of the
SDSS galaxies, given the relatively large projected aperture size
of the fibers (5.5 kpc diameter at $z=0.1$).   
This is much larger than the $\sim$200 pc apertures used in the
survey of nearby galaxy nuclei by Ho, Filippenko \& Sargent (1997). It is
therefore not
surprising that the majority of AGN in our sample fall into the ``transition''
class, and have line ratios intermediate between those of
star-forming galaxies and those of LINERs or  Seyferts.

We therefore prefer an AGN classification system that is less sensitive to
aperture, while still reflecting the difference in intrinsic nuclear luminosity
between Seyfert galaxies and LINERs. Because we are studying type 2 systems, we seek an AGN
component that emits radiation isotropically and that can be assumed to provide some
indication of the total level of nuclear activity in the galaxy. In the optical,
it is traditional to consider the luminosity of the narrow-line
emission region (NLR), which, in principle, should not be strongly affected by the
obscuring torus that surrounds the central source. 

In this paper, we focus on the luminosity of
the [OIII]$\lambda$5007 line as a tracer of AGN activity. Although this line can be excited by
massive stars as well as an AGN,
it is  known to be relatively 
weak in metal-rich, star-forming galaxies. This can be seen in Fig. 1:  
star-forming galaxies define a sequence in which 
the ratio [OIII]/H$\beta$ decreases and the ratio [NII]/H$\alpha$ increases 
towards higher gas-phase  metallicities (e.g. Charlot \& Longhetti 2001).
The AGN sequence emerges as a plume from the bottom of the
locus of star-forming galaxies. The morphology of the plume is strongly
suggestive of  a ``mixing line'', in which
the relative contribution of the AGN increases from lower left to upper right 
(see Kewley et al (2001) for a discussion).
Fig. 1 also suggests  that the ``contaminating'' emission is from
metal-rich, star-forming regions, because the AGN sequence emerges
from the bottom of the locus of star-forming galaxies. As we
show later, this is expected because AGN  are located in
massive galaxies. Star-forming galaxies
are known to exhibit a strong mass-metallicity relation (Tremonti et al. 2003).
If AGN hosts lie on this relation, their metallicities will be solar or higher,
and the contamination of the [OIII] line by star-formation
will be small.

The [OIII] line also
has the advantage of being strong and easy to detect in most
galaxies. It should be noted that because the  narrow-line emission arises 
outside the dust sublimation radius, it is affected by
dust within the host galaxy. Thus it is important to correct our [OIII]
luminosities
for the effects of extinction. We can measure the extinction using the
Balmer decrement. This procedure has clear physical meaning in the 
``pure'' Seyfert 2's and LINERs. In the case of the transition
objects, the lines will arise both in the NLR and the surrounding HII
regions, with a greater relative AGN contribution to [OIII] than to the Balmer
lines. Thus,
a dust correction to [OIII] based on the ratio H$\alpha$/H$\beta$  
should  be regarded as at best approximate.

In Fig. 2  we plot the [OIII]/H$\beta$ versus [NII]/H$\alpha$ BPT diagram in
bins of
extinction-corrected [OIII] line luminosity. Fig. 2 shows clearly that  
the region of the BPT diagram occupied by galaxies classified as  LINERs
is primarily populated by objects with
low [OIII] line luminosities. Conversely, the region of the diagram
with [OIII]/H$\beta> 3$ and  [NII]/H$\alpha > 0.6$ (i.e. the Seyfert
region of the diagram)  
is mainly populated with galaxies that have high [OIII] luminosities.
This demonstrates that classification by line ratio and classification
by [OIII] luminosity are roughly equivalent for LINERs and Seyferts.

The virtue of classifying galaxies by [OIII] luminosity is that this then
allows us to study the large 
number of transition galaxies with spectra intermediate between pure 
star-forming systems and pure LINERs/Seyferts. 
Fig. 2 shows that transition galaxies span a wide range in [OIII]
luminosity, but have very similar line ratios. This means that one cannot
distinguish between transition objects containing a Seyfert 2 nucleus and those
containing  a LINER
on the basis of position in the BPT diagram alone. Adding the extra
dimension of [OIII] luminosity, the separation is clear.
We  stress that any attempt to characterize the stellar
population in  AGN hosts  must  include
the transition objects that comprise the majority of the AGN in
Fig. 1. Excluding these would bias the sample against host galaxies
with significant amounts of on-going star-formation. 

\begin{figure}
\centerline{
\epsfxsize=15cm \epsfbox{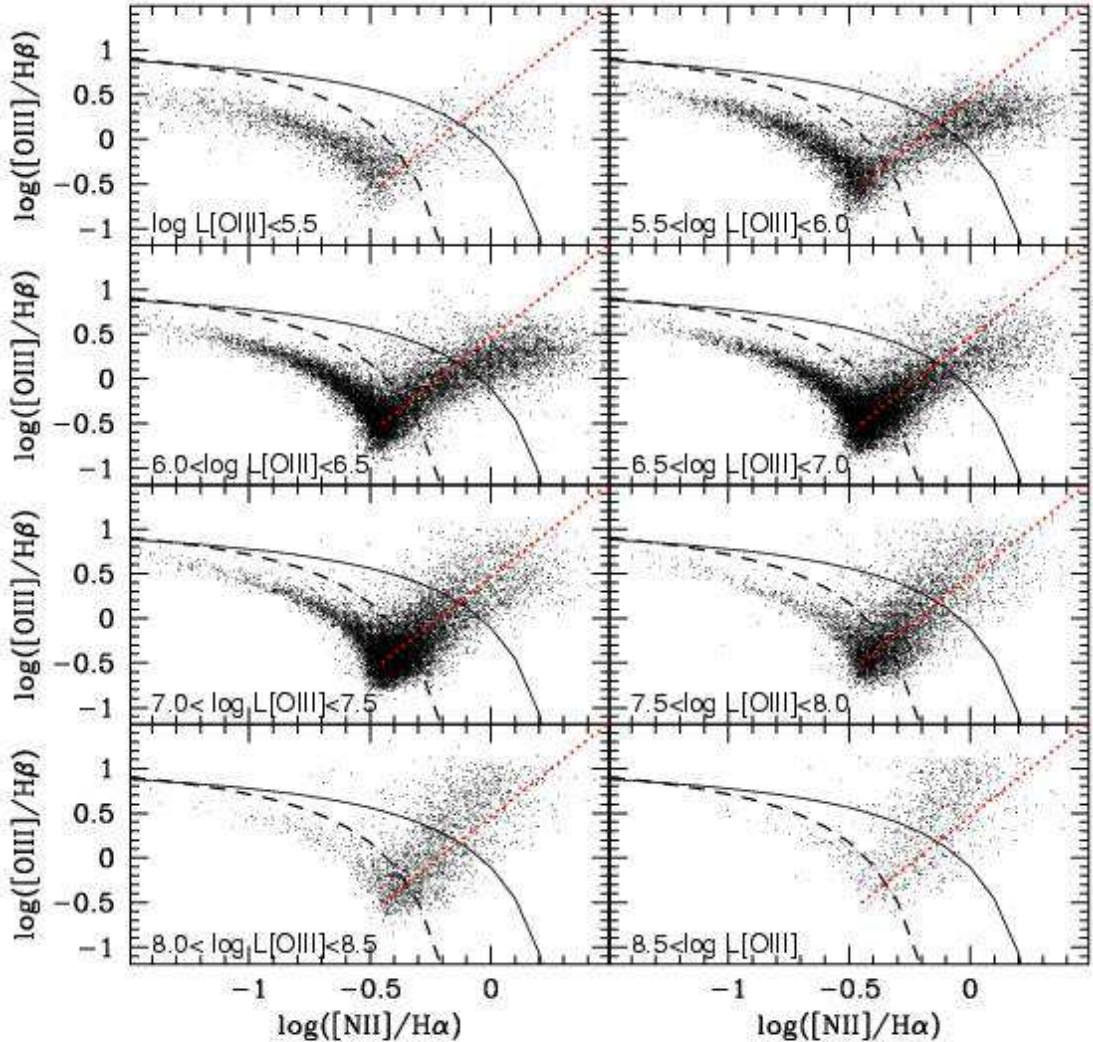}
}
\caption{\label{fig2}
\small
The BPT diagram of Fig. 1 has been binned into fixed ranges in extinction-corrected
[OIII] line luminosity. Note that luminosities are given in units of the
bolometric solar luminosity ($3.826 \times 10^{33}$ erg s$^{-1}$). The dashed black curve
shows our adopted demarcation between star-forming galaxies and AGN. The solid curve
shows the Kewley et al demarcation. 
The dotted red lines mark          
the angle $\Phi=25^{\circ}$, which separates pure LINERs from 
Seyferts  reasonably
cleanly.}
\end {figure}
\normalsize

We propose the following simple scheme for characterizing the position of
an emission line galaxy on the BPT diagram of Fig. 1. We define an ``origin''
{\bf O} located at [NII]/H$\alpha$=-0.45 and [OIII]/H$\beta$=-0.5, i.e.
near the bottom of the locus of star-forming galaxies where it                    
intersects the lower end of the AGN sequence.
We then parametrize the position of each galaxy by its distance $D$
from the origin {\bf O}, and  by an angle $\Phi$, which is defined to be
zero in the direction parallel to the positive [OIII]/H$\beta$ axis,
and which increases as the galaxy moves in a clockwise direction towards 
increasing values of [NII]/H$\alpha$ and decreasing values of [OIII]/H$\beta$.
In this scheme, pure Seyferts are characterized by large values of $D$ and small
values of $\Phi$ ($0^{\circ}-25^{\circ}$), whereas ``pure'' LINERs are
characterized by large values of $D$ and  values of $\Phi$ in the range
$25^{\circ}-60^{\circ}$. Transition objects are characterized by
small values of $D$.

The left panel of Fig. 3 shows the correlation between [OIII] line luminosity and
$\Phi$ for the AGN in our sample. In this plot, we have
excluded the transition objects by selecting  only those AGN  
that lie above the Kewley et al (2001) demarcation
curve. There is a strong transition in the median AGN luminosity at
$\Phi \sim 25^{\circ}$. 
In the right panel of Fig. 3, we plot L[OIII] as a function of $D$.
There is no strong trend              
in intrinsic  AGN luminosity as a function of distance from the locus of
star-forming galaxies. This supports the idea that AGN                  
define two basic ``mixing sequences'':
a star-formation plus high-luminosity
AGN (Seyfert 2 sequence) and star-formation plus low-luminosity AGN
(LINER sequence).  

\begin{figure}
\centerline{
\epsfxsize=10cm \epsfbox{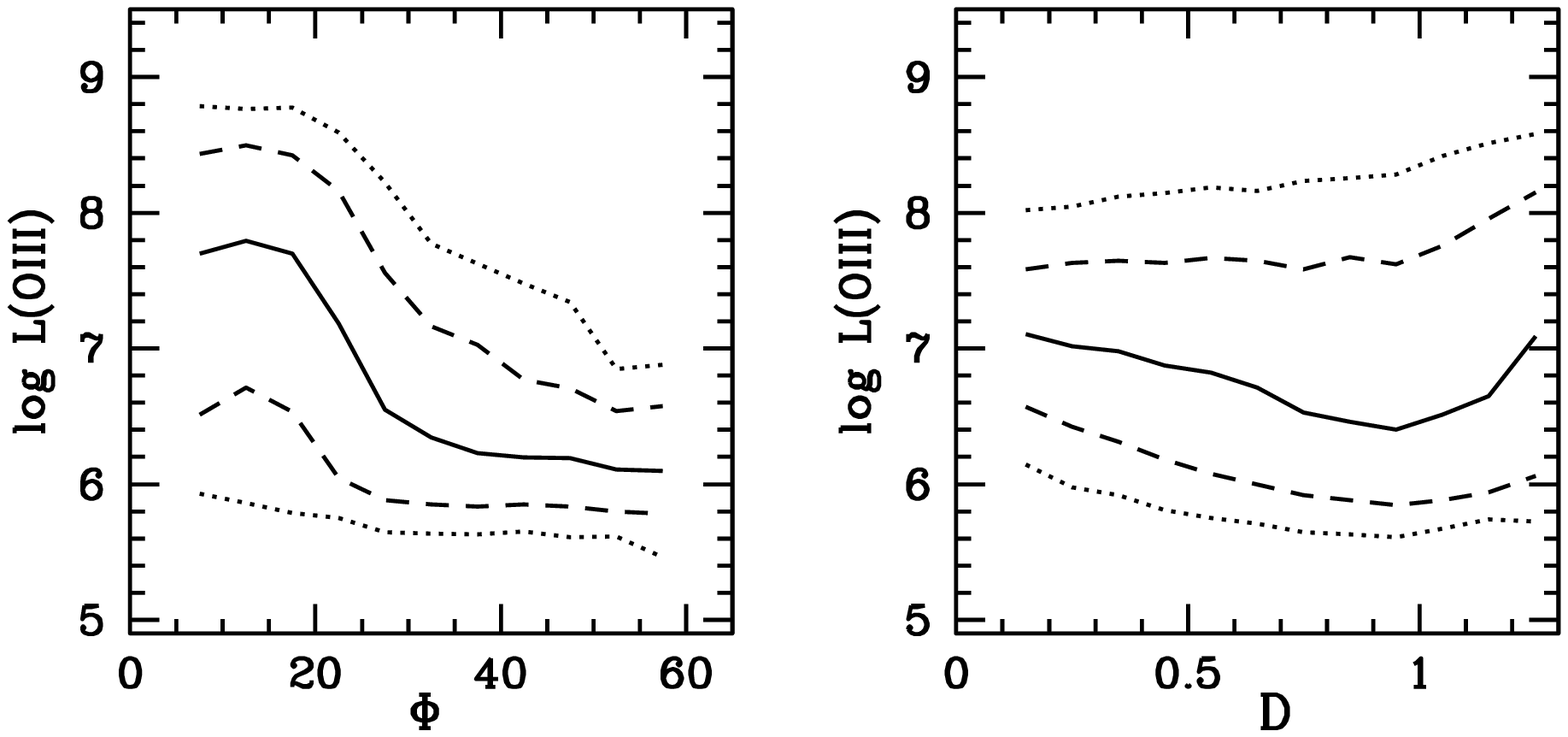}
}
\caption{\label{fig3}
\small
Left: The [OIII] line luminosity is plotted as a function of the angle parameter $\Phi$
for all AGN lying above the Kewley et al. demarcation curve in Fig. 1.
Right: The [OIII] line luminosity is plotted as a function of $D$, the distance from the locus
of star-forming galaxies.
In each plot the solid line shows the median relation as a function of
$D$ or $\Phi$, while the dashed and
dotted lines indicate the 16-84 and 2.5-97.5 percentiles in $\log$ L[OIII]   
respectively.}
\end {figure}
\normalsize

\subsection {The Effect of the AGN on the Derived Host Galaxy Properties}

The traditional view of type 2 AGN is that their spectra are composed of an old, metal-rich
stellar population plus
an underlying ``featureless continuum'' produced by the central source (e.g. Koski 1978). This view has been
challenged in recent years by a number of studies showing that the spectra are, in fact, better fit
by stellar populations that span a wide range in age and metallicity.

Schmitt, Storchi-Bergmann \& Cid Fernandes (1999) have carried out a detailed spectral synthesis
analysis of spectra of the nuclear regions of 20 nearby Seyfert 2 galaxies. 
They  fit  each observed spectrum  using  a grid of 
equivalent widths and continuum ratios measured from star clusters
of different ages and metallicities. In addition they include  a $F_{\nu} \propto \nu^{-1.5}$
power-law component to represent a canonical AGN continuum.
The code  considers {\em all possible} linear combinations of age, metallicity and
featureless continuum, weighting each solution by $e^{-\chi^2/2}$ in order to construct
the likelihood function of parameters such as stellar age or fractional  contribution of the 
AGN continuum.
The main result of this analysis is that the continuum contribution of the AGN is extremely small, 
rarely exceeding 5 percent.
The analysis in  Schmitt et al (1999)  focused on the nuclear stellar populations of type 2
Seyferts. As discussed in section 2, the SDSS spectra include much more galaxy
light, so we expect
the contribution from the featureless continuum to be even smaller.

It is important to establish that this is the case.
Throughout this paper we will use the strength of the 4000 \AA\ break
as our primary indicator of the age of the stellar population. Any  contribution
of AGN light would decrease the strength of the 4000 \AA\ break,
causing us to underestimate the age of the stellar
population. It should be noted, however, that stellar Balmer absorption-lines become weaker
with age, and for ``well-behaved'' (continuous)
past star-formation histories, the equivalent widths of the Balmer lines are inversely correlated with 
the 4000 \AA\ break strength (see
Paper I). Thus, the addition of AGN light to the Balmer lines would cause us to {\em overestimate}
the stellar age. A clear test
of the importance of AGN light is to compare the Balmer
absorption lines in AGN hosts with the Balmer lines in  normal galaxies that have the same 4000 \AA\ break
strength. If AGN light is significant, the Balmer lines will be 
{\em weaker} in the AGN hosts.

To perform this test,             
we have combined the spectra of our highest [OIII] luminosity AGN  to create a high
signal-to-noise {\em composite} spectrum. This is plotted in black in Fig. 4. For each AGN
included in the composite, we picked out a random galaxy from the parent sample with the same stellar mass, 
4000 \AA\ break strength and redshift as the AGN host. We combined the spectra of these ``matching'' galaxies
to create the spectrum shown in red. As can be seen, the AGN spectrum and the galaxy
spectrum are identical except for the emission lines, which are
stronger in the AGN (as expected). In particular, 
the high-order Balmer absorption-lines match extremely well.
We have also experimented with fitting the AGN template spectrum with a combination of BC2003 
model spectra and a canonical QSO spectrum (Vanden Berk et al 2001). We find that we cannot obtain
an acceptable fit if the QSO contributes more than a few percent of the light
in the $r$-band.

\begin{figure}
\centerline{
\epsfxsize=15cm \epsfbox{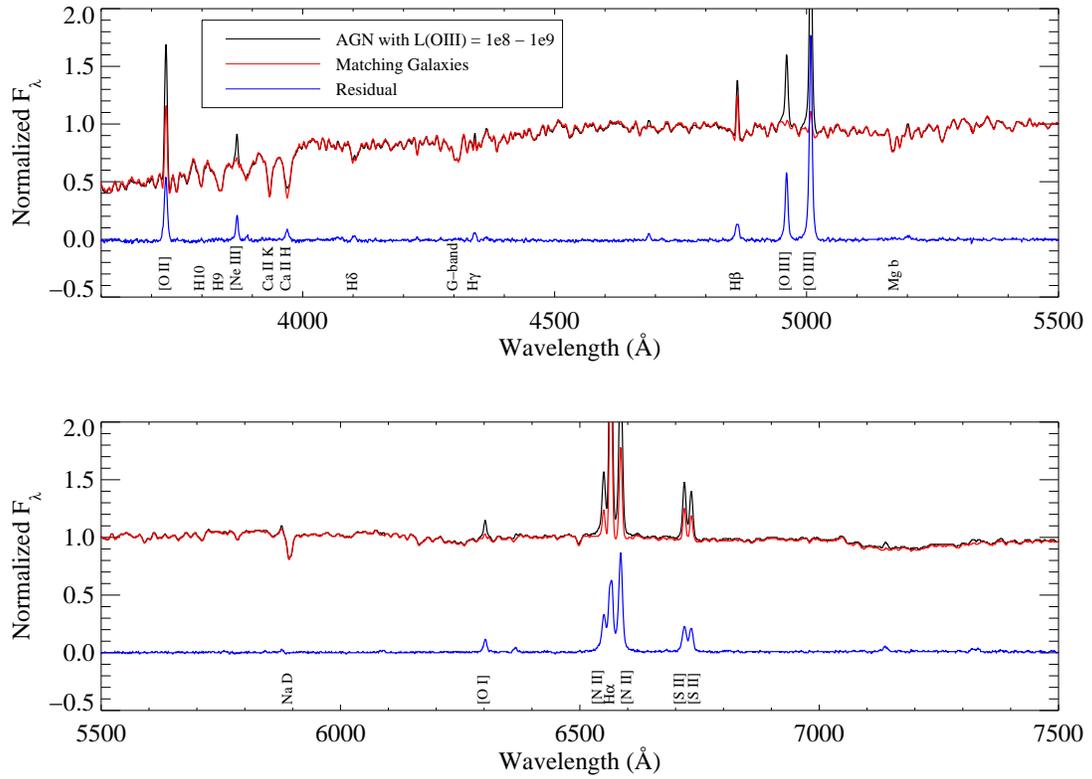}
}
\caption{\label{fig4}
\small
A comparison of the stacked spectra of 500 AGN with [OIII] line luminosities in the range $10^8-10^9$ solar         
(black) with the stacked spectra of a set of matching galaxies (red). The blue line shows the
residual when the galaxy spectrum is subtracted from the AGN spectrum.}
\end {figure}
\normalsize

It is also important to consider the effect of the AGN emission on the basic galaxy
structural properties derived from the SDSS imaging data. In particular,
we have used the $z$-band absolute magnitude to calculate the host
galaxy stellar mass ($M_*$) and surface mass density ($\mu_*$), and
have used the $r$-band images to measure the concentration parameter
($C$). From Fig. 4, it is clear  that the continuum
in AGN hosts is dominated by starlight and that any AGN continuum emission
will not affect our derived parameters.
The effect of AGN emission-lines will likewise be small.
The $z$-band
samples a spectral region in the rest-frame of a typical AGN host
that is devoid of strong emission-lines. For $z \leq$ 0.06 the
[NII]$\lambda$$\lambda$6548,6584 and H$\alpha$ emission-lines fall
in the $r$ bandpass, while the [OIII]$\lambda$$\lambda$4959,5007 lines move
in at $z \geq$ 0.11. Fortunately, either set of lines has a combined equivalent
width ranging from $\sim$10 \AA\ in weak AGN to $\sim$100 \AA\
in powerful
AGN. Since the SDSS $r$ bandpass is $\sim$1400 \AA\ wide, the flux
contribution of the emission-lines to the $r$-band flux
will be at most a few percent in the inner 3 arcsec region of the galaxy.

In this section, we have argued that AGN light does not contribute significantly 
to the continuum emission in type 2 AGN.                       
Regions of star-formation do, however, 
make a significant contribution to the emission-line spectra
of the AGN in our sample, because of the large projected
aperture size of the SDSS fibers. 
Because the two composite spectra plotted in Fig. 4 are matched in
stellar mass and in mean stellar age, it is reasonable to assume that the galaxies 
included in the two samples have similar amounts of on-going star-formation. Comparison
of the emission-line properties of the two spectra demonstrates 
that in the AGN composite spectrum, star-forming regions contribute 50\% of
the [OII]$\lambda$3727 flux, 70\% of the H$\beta$ flux, 65\% of the H$\alpha$
flux, and 45\%  of the [NII]$\lambda$6584 and
[SII]$\lambda$$\lambda$6717,6731 flux. In contrast, star-forming regions contribute only
7\% of the [OIII]$\lambda$5007 flux. The [OIII]$\lambda$5007
line is clearly a far better measure of the AGN power than any of the other
strong emission-lines. Croom et al. (2002)
reach qualitatively similar conclusions for low-redshift QSOs.

\section {The Host Galaxies of Type 2 AGN}                                                            

In this section we study the properties of the host galaxies of the AGN in our
sample, including their stellar masses, their surface mass densities and concentrations,
and their mean stellar ages as measured by spectral indices such as the
4000 \AA\ break strength.

Note that the AGN in our sample are drawn from an $r$-band limited spectroscopic survey of galaxies.
Galaxies of different luminosities can be seen to different distances
before dropping out due to the selection limits of the survey.
The volume $V_{max}$  within which a galaxy can be seen and will be included in the sample
goes as the distance limit cubed, which results in the     
samples being dominated by intrinsically bright galaxies.
Throughout this analysis, we will correct for this effect by giving each galaxy or AGN
a weight equal to the inverse of its maximum
visibility volume determined from the apparent magnitude limit of the survey
(see Paper II for more details). All correlations we plot should then be appropriate
for a volume-limited sample of galaxies and should no longer be affected by selection biases
due to the criteria which define the main galaxy sample.

\subsection {Stellar Mass and Aperture Effects}  

As discussed in section 3, our sample of emission-line galaxies includes 55,747 objects where
the four emission lines, H$\alpha$, [NII], [OIII] and H$\beta$,
are all detected with $S/N >3$. In total, forty percent of these galaxies
are classified as AGN, but this number is strongly dependent on host mass and to
a lesser extent, on morphological type.
This is illustrated in  Fig. 5. 
In the top panels, the blue histogram shows the fraction of all  galaxies that have emission lines  
as a function of $\log M_*$ and $C$.
The red histogram shows 
the fraction of all galaxies that are classified as AGN.
The black histogram shows the fraction of all {\em emission line galaxies}
that are classified as AGN.
As can be seen, AGN are preferentially found in more massive and more
concentrated galaxies. The dependence  of AGN fraction on mass
is very  striking. Even though 70-80\% of galaxies
with stellar masses less than $10^{10} M_{\odot}$ have detectable emission lines,
only a few percent are classified as AGN. In contrast, more than
80\% of emission line galaxies with $M_* > 10^{11} M_{\odot}$
are AGN.  

It should be noted that AGN detection rates are subject to strong selection effects.
Ho et al (1997) find that
43\% of the galaxies in their survey can be regarded as active and nearly all
their galaxies have detectable emission lines.
In our sample, the nuclear spectrum will be  
diluted by the light from  surrounding host galaxy, so our derived  AGN
fractions are lower.
In the middle left panel of Fig. 5  we plot
the AGN fraction as a function of  normalized distance $z/z_{max}$, where
$z_{max}$ is the maximum redshift out to which the galaxy would have been included
in the SDSS main galaxy sample. 
The red line shows the AGN fraction for ``massive'' galaxies with stellar
masses in the range $3\times 10^{10} - 10^{11} M_{\odot}$. The blue line
is for low mass galaxies in the range $10^8 - 3\times 10^9 M_{\odot}$.
The AGN fraction in massive galaxies is a very strong function
of distance. At low values of $z/z_{max}$, the AGN fraction  reaches 
$50$ \%, a value very similar to the fraction that  Ho et al found for
the $L_*$ galaxies in their sample.
On the other hand, the AGN fraction in low mass galaxies does not rise above
a few percent, even at low values of $z/z_{max}$. There are  6586 galaxies in our sample
with stellar masses in the range $10^8-3\times 10^9 M_{\odot}$, so this result has high
statistical significance.  

We now ask whether it is possible to define an AGN fraction  that does not
depend on aperture. In the middle right panel of Fig. 5, we plot
the fraction of AGN in galaxies with
stellar masses in the mass range $3 \times 10^{10} -10^{11} M_{\odot}$ and with [OIII] 
line luminosities greater than $3 \times 10^6$ (red),
$10^7$ (green) and $3 \times 10^7$ (blue) $L_{\odot}$ as a function of normalized distance.
For [OIII] luminosities greater than $10^7$ $L_{\odot}$ , 
the AGN fraction no longer depends on $z/z_{max}$.
We will henceforth refer to AGN as ``strong'' or ``weak'' using
L[OIII] = $10^7 L_{\odot}$ as a dividing line. Fig. 5 shows
that our sample of strong AGN is complete. Figs 2 and 3 imply
that the division very roughly corresponds to that between
Seyferts and LINERs. Finally, in the bottom two panels of Fig. 5 we plot
the fraction of ``strong'' AGN in galaxies as a function of $\log M_*$ and $C$.  

\begin{figure}
\centerline{
\epsfxsize=13cm \epsfbox{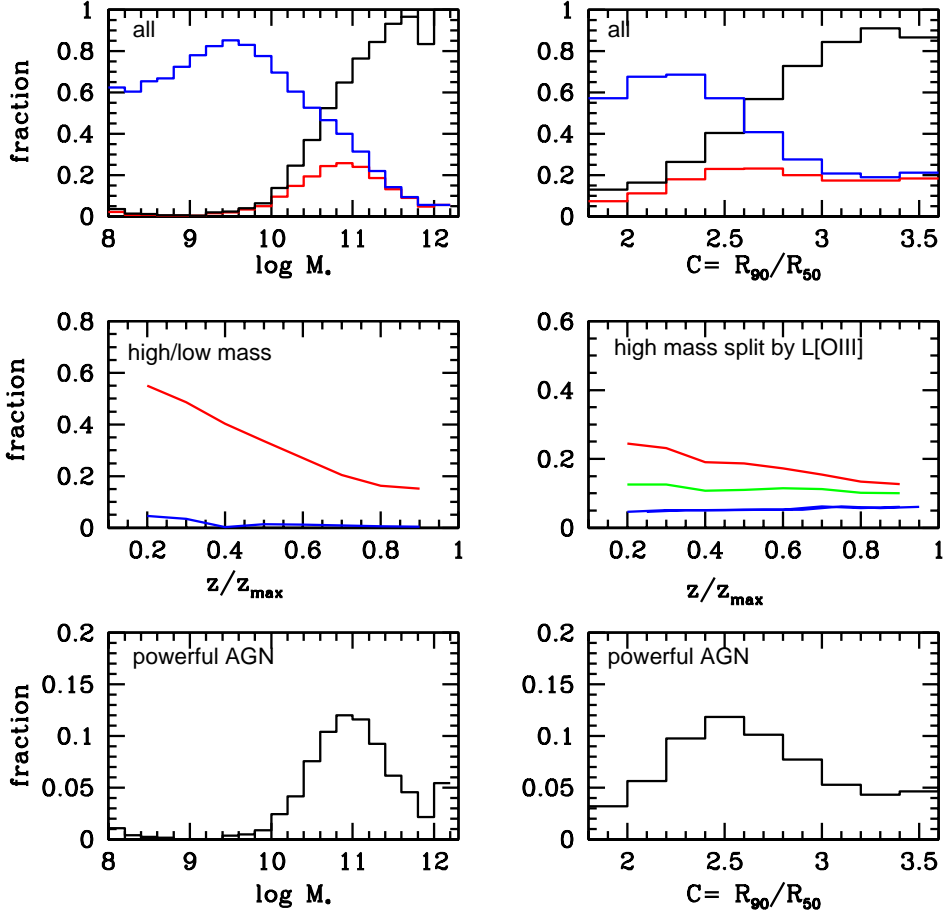}
}
\caption{\label{fig5}
\small
{\bf Top panels}:  The blue histogram shows the fraction
of all galaxies in our sample  with $S/N>3$ detections of 
[NII], H$\alpha$, [OIII] and H$\beta$ as a function
of $\log M_*$ and concentration index $C$. The red histogram shows the fraction of all galaxies classified as AGN.
The black histogram shows the fraction of emission-line galaxies classified
as AGN. {\bf Middle left:} The fraction of galaxies with $3\times10^{10} < M_* < 10^{11} M_{\odot}$
classified as AGN is plotted as a function of $z/z_{max}$ as a red line . The blue line is
for galaxies with $3\times 10^{8} < M_* < 10^{9} M_{\odot}$. {\bf Middle right:}
The AGN fraction for  galaxies with $3 \times 10^{10} < M_* < 10^{11} M_{\odot}$
and with $\log L[OIII] >$ 6.5 (red), 7(green) and 7.5(blue) is plotted
against $z/z_{max}$. {\bf Bottom panel:} The fraction of galaxies containing AGN with
$\log L[OIII] >$ 7 is plotted as a function of $\log M_*$ and $C$.}
\end {figure}
\normalsize

The most striking result in Fig. 5 is that very few AGN are found in galaxies
with $M_* < 10^{10} M_{\odot}$. As discussed in Paper II, the majority of low
mass galaxies have young stellar populations and a significant fraction are
experiencing ``bursts'' of star formation at the present day.
Many low mass galaxies thus have strong emission lines 
due to star formation and one might therefore be concerned whether our BPT diagnostic
method would be able to detect an AGN in such objects.
In the top left panel of Fig. 6, we plot the
[OIII]/H$\beta$ versus [NII]/H$\alpha$ BPT diagram for  
emission-line galaxies that are not classified as AGN and that
have  stellar masses in the range $10^8 - 3 \times 10^{9} M_{\odot}$.
The next 3 panels in Fig. 6 illustrate  the effect of adding successively more
powerful AGN to these galaxies.   
We have extracted ``pure'' AGN lying above
the Kewley et al demarcation curve in Fig. 1 and we have ranked these AGN according
to their extinction-corrected  [OIII] luminosities. For each low mass  
galaxy, we select a random  AGN 
and add its observed emission-line  luminosities to those
of the original galaxy. Fig. 6 shows that  the presence of even a low-luminosity AGN with
$10^5 <$ log L[OIII] $< 10^6 L_{\odot}$ perturbs the emission line ratios
enough to allow the AGN to be identified in 93\% of the galaxies in the sample.
The bottom right panel of Fig. 6 demonstrates that a  powerful AGN 
with L[OIII] $ > 10^7 L_{\odot}$ 
would be detected in more than 99\% of the objects.

Although we cannot rule out the presence of
extremely weak active nuclei in low mass galaxies, such as the AGN with L[OIII]=
$5 \times 10^4$ $M_{\odot}$  discussed by Kraemer et al (1999)
in the Sdm galaxy NGC 4395,
we conclude that more powerful AGN
are confined almost exclusively to high mass galaxies. Ulvestad \& Ho (2002)
have recently reached similar conclusions using radio observations of a sample of 40
late-type galaxies.
 
\begin{figure}
\centerline{
\epsfxsize=12cm \epsfbox{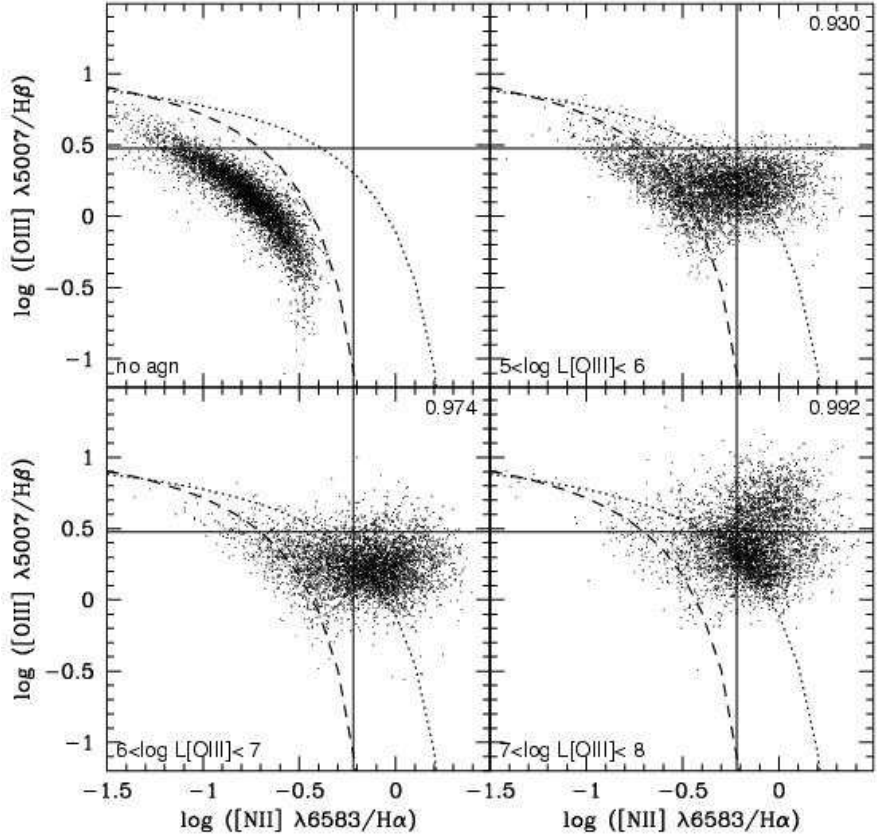}
}
\caption{\label{fig6}
\small
{\bf Top left}:  The BPT diagram of Fig.1 for galaxies with $8 < \log M_* < 9.5$
that are not classified as AGN. {\bf Next 3 panels:} The effect of adding
AGN of increasing [OIII] luminosity (see text for more details).
The numbers in the upper right corners of each panel indicate
the fraction of galaxies that lie above the dashed line.}
\end {figure}
\normalsize

There are many studies in the literature that
indicate a positive correlation between AGN power and host galaxy 
luminosity or linewidth.  
Most of these studies have been restricted 
to specific classes of AGN, for example Seyferts (e.g. Nelson \& Whittle 1996)
QSOs (e.g. Kotilainen \& Ward 1994; Croom et al. 2002), and radio galaxies
(e.g. Ledlow \& Owen 1996).  In the left panel of Fig. 7 we plot
as a solid line the median stellar mass $M_*$ of AGN host galaxies
as a function of   [OIII] luminosity. 
The dashed lines indicate the 16-84  percentiles
of the $M_*$  distribution at a  given value of L[OIII].  
Fig. 7 shows that galaxies of given  mass are able to host
AGN that span a very wide (more than several orders of magnitude) range in [OIII] luminosity.   
Note that we are unable to detect AGN with [OIII] luminosities much below
$10^4 L_{\odot}$, so the actual range may be considerably larger. 
The median stellar mass of the host depends only weakly on [OIII] luminosity,
but we do find that more powerful AGN are located preferentially in more massive host galaxies.

In the right panel of Fig. 7, we plot
the {\em maximum} [OIII] luminosity that is attained by galaxies of a given stellar mass.
In order to estimate L[OIII](max) in an
unbiased way, we have ordered all 122,808 galaxies in our sample according to
stellar mass. We then partition the sample into mass bins that contain an equal
number of galaxies (2000 in this case).
The dotted line shows the upper 99th percentile of the [OIII]
luminosity distribution  in each mass bin. 
(Note that the most [OIII]-luminous  galaxies are indeed AGN for  $M_* > 10^{10} M_{\odot}$,  
but are mostly star-forming galaxies for $M_* < 10^{10} M_{\odot}$.)
The solid line shows the upper 97.5 percentiles of the distribution.
 As can be seen, L[OIII](max) scales roughly  linearly
with stellar mass at $10^{10} < M_* < 10^{11} M_{\odot}$. At high stellar masses
L[OIII](max) flattens and there is even evidence for a turnover
at the highest masses ($M_* > 3 \times  10^{11} M_{\odot}$).

\begin{figure}
\centerline{
\epsfxsize=12cm \epsfbox{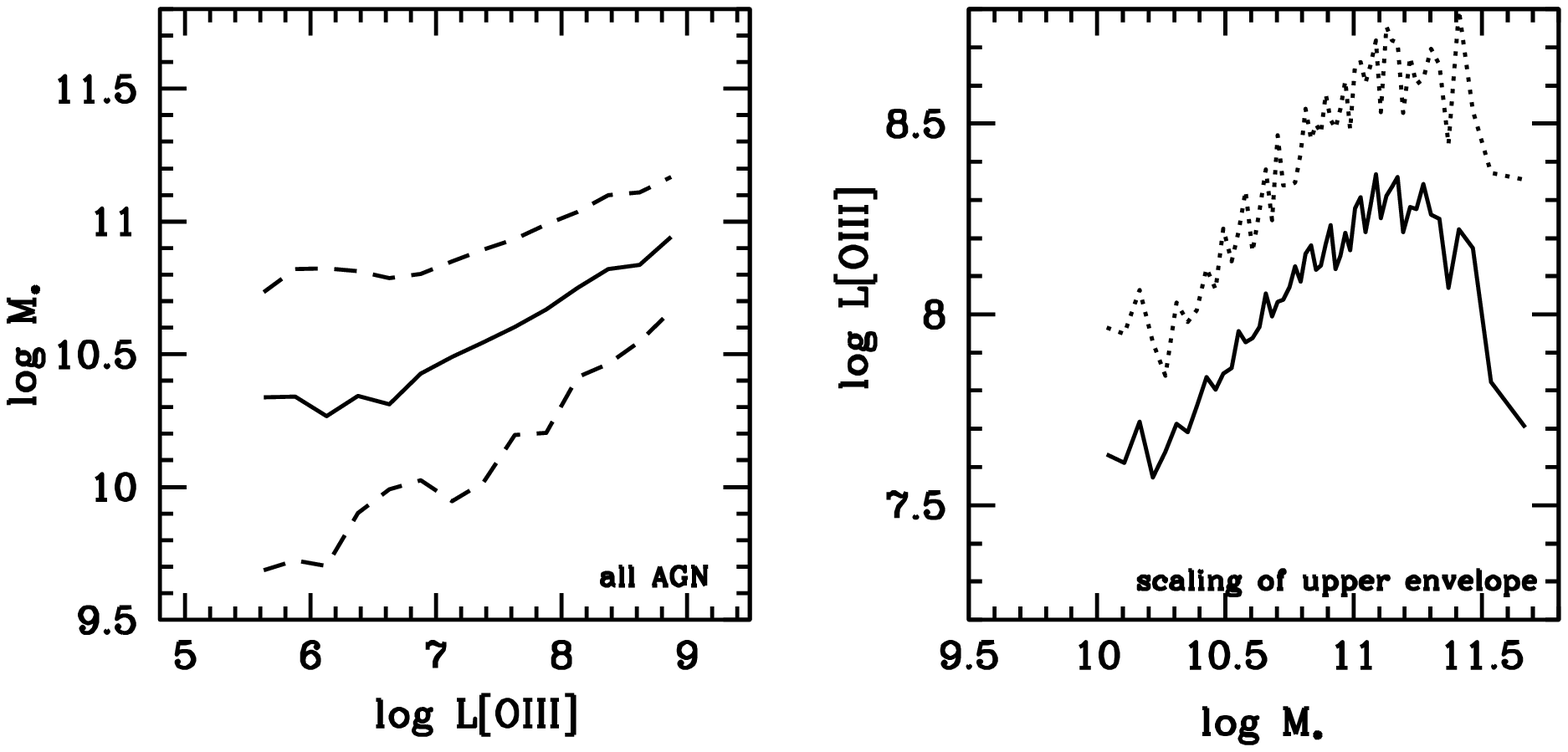}
}
\caption{\label{fig7}
\small
{\bf Left:} The distribution of stellar mass as a function of  [OIII] luminosity 
for the AGN in our sample. The solid line shows the median value of L[OIII], while
the dashed lines show the 16-84  percentiles of the
$1/V_{max}$ weighted distribution. {\bf Right:} The solid line shows the upper 97.5 percentile of
the [OIII] distribution of all galaxies as a function of stellar mass. The dotted line is the
upper 99 percentile.}
\end {figure}
\normalsize

\subsection {Surface Mass Density and Concentration}

Most previous studies of the global morphologies 
of type 2 AGN hosts have concentrated on  
Hubble type. Early    
morphological studies of Seyfert galaxies by Adams (1977) and Heckman (1978) revealed that
most Seyferts are in spiral galaxies, primarily early-type spirals (Sa-Sb).
AGN in the survey of nearby galaxy nuclei by  Ho, Filippenko \& Sargent (1997)
are found in  ellipticals, in lenticulars and in bulge-dominated spirals, with
LINERs occupying earlier-type hosts than Seyferts or transition systems.

Because it is impractical to classify hundreds of thousands of galaxies by eye,
studies of galaxy morphology  in the SDSS have focused on                       
simple structural parameters that can be measured automatically for
very large numbers of objects. In section 2, we defined two  such parameters: 
the concentration index $C$ and the stellar surface mass density $\mu_*$.
As discussed in Strateva et al (2001) and in  Paper II ,
early-type galaxies (Hubble type Sa,S0 and E)  have $\mu_*$ in the range $3 \times 10^8 - 
3 \times 10^9 M_{\odot}$ kpc$^{-2}$ and $C>2.6$.
Late type galaxies have 
$\mu_* < 3\times 10^8 M_{\odot}$ kpc$^{-2}$ and $C<2.6$.

In Fig. 8 we plot the $1/V_{max}$ weighted distributions of 
$\mu_*$ and $C$ as a function of [OIII] luminosity for 
the AGN in our sample. 
AGN of all luminosities are found  in galaxies with 
high surface densities and intermediate concentrations. 

\begin{figure}
\centerline{
\epsfxsize=12cm \epsfbox{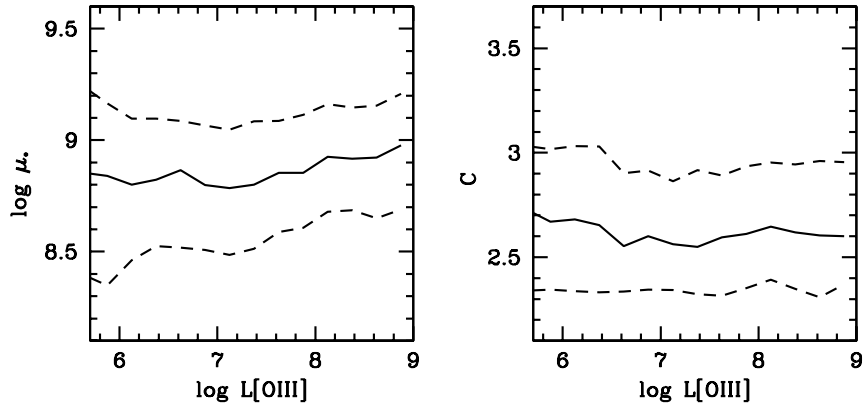}
}
\caption{\label{fig8}
\small
The stellar surface density $\mu_*$ and concentration index $C$ are plotted  
as functions of log L[OIII] for the AGN in our sample. The solid
line shows the median, while the dashed lines indicate the 16-84 percentiles of the
$1/V_{max}$ weighted  distribution.}
\end {figure}
\normalsize

In Paper II, we showed that  $C$ and $\mu_*$ correlate strongly with stellar mass in
ordinary galaxies.
A sharp transition in both $M_*$ and $C$
takes  place at a characteristic stellar mass of $\sim 3 \times 10^{10} M_{\odot}$.
If the sample is split into early- and late-type subsamples at a $C$ index
value of 2.6, the surface densities of late-type galaxies 
scale with stellar mass as $\mu_* \propto M_*^{5/3}$. 
The surface densities of early-type galaxies, on the other hand, 
are independent of stellar mass, and have  a median value of
$10^9 M_{\odot}$ kpc$^{-2}$.

Fig. 9 compares the $\mu_*-M_*$ and $C-M_*$ relations for AGN  with those of  
ordinary galaxies. The grey-scale indicates the fraction of galaxies in a given
logarithmic mass bin that fall into each bin in surface density
or concentration (see Paper II for more details on how these
conditional density distributions are computed). Note that we only plot the relations    
in mass bins that contain more than 100 objects. This is why the relations
cut off below $10^{10} (10^{9.5}) M_{\odot}$ for the strong (weak) AGN.
Fig. 9 shows that strong AGN (defined to have log L[OIII]$ > 7.0$)
show very little dependence of either concentration or surface
mass density on stellar mass. At all masses, they have the high densities
typical of early-type galaxies and concentrations that are intermediate between
early and late-type galaxies.
Both the concentrations and the surface densities of the weak AGN (log L[OIII]$<7.0$) 
are similar to those of ordinary galaxies of the same stellar mass.

\begin{figure}
\centerline{
\epsfxsize=15cm \epsfbox{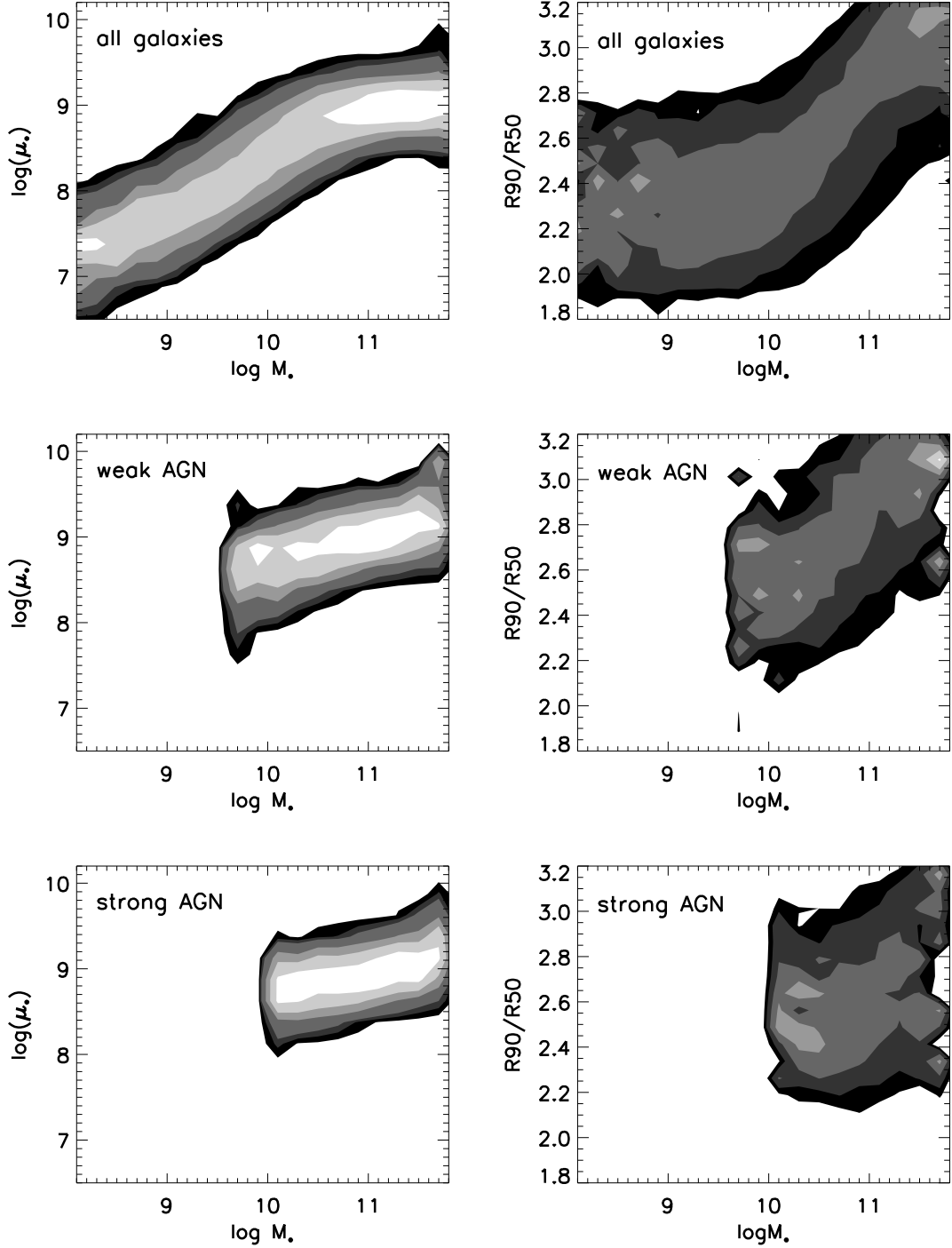}
}
\caption{\label{fig9}
\small
Left: Conditional density distributions showing trends in the surface mass density $\mu_*$ as a function
of stellar mass $M_*$ for all galaxies (top), for weak AGN with log L[OIII]$ < 7.0$
(middle) and for strong AGN with log L[OIII]$ > 7.0$ (bottom).
Each object has been weighted by $1/V_{max}$ and the bivariate distribution has been
normalized to a fixed number of galaxies in each bin of $\log M_*$.
Right: Conditional density distributions showing trends in concentration index $C$ as a function of
$\log M_*$. In this and all other conditional density plots, each contour
represents a factor 2 increase in density.}
\end {figure}
\normalsize

In Fig. 10, we compare the surface mass density distributions of AGN hosts to those
of ordinary early-type and late-type galaxies.
We have separated early-type galaxies from late-types  at a $C$ index of 2.6, and
we show results for four independent ranges in $\log M_*$.
Fig. 10 shows that the surface density distributions of AGN host galaxies strongly resemble
those of early-type galaxies of the same stellar mass.

\begin{figure}
\centerline{
\epsfxsize=14cm \epsfbox{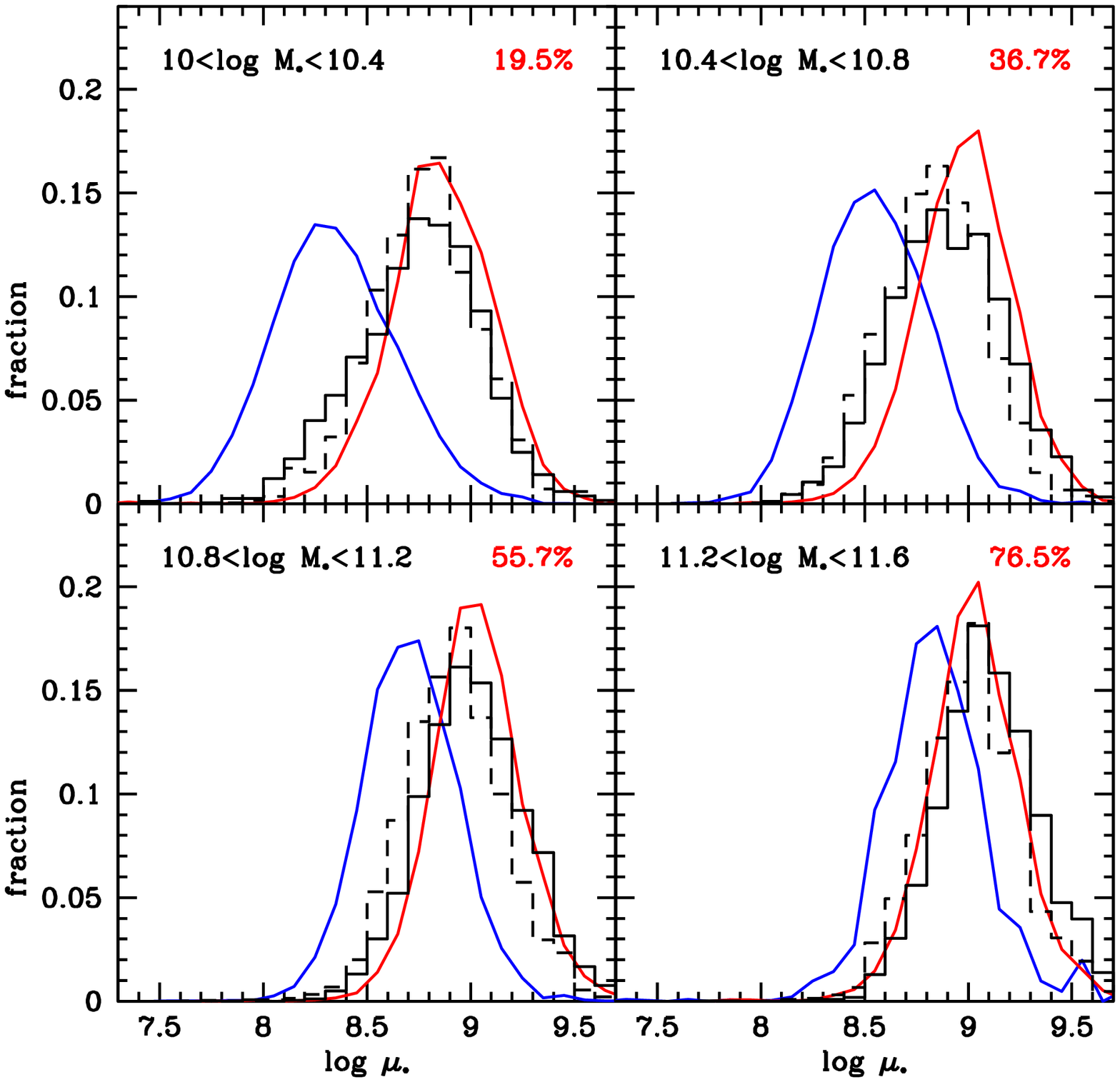}
}
\caption{\label{fig10}
\small
The surface mass density distributions of the host galaxies of weak AGN (solid black) and
of strong AGN (dashed black) are compared to
those of early-type ($C>2.6$) galaxies [red] and late-type ($C<2.6$) galaxies [blue]
in 4 separate ranges of $\log M_*$. The numbers listed in the top
right corner of each panel indicate the percentage of galaxies   
in that mass range that are early-type.}
\end {figure}
\normalsize

Finally, Fig. 11 shows how the surface densities and concentrations of AGN hosts
vary as a function of position on the BPT diagram.
AGN that lie close to the sequence  of star-forming galaxies in the BPT diagram
occur in lower-density, less concentrated  hosts. This is consistent with
the idea that the transition objects occur in galaxies 
of later type than ``pure''
Seyferts or LINERs.
On the other hand, there appears to be rather little trend in the  $\mu_*-M_*$ or $C-M_*$ 
relations as a function of the ionization-sensitive angle parameter $\Phi$ (Note that in
the $\Phi$ plots 
we have restricted the sample to objects lying above the Kewley et al (2001) line,
whereas in the $D$ plots we use all objects above our canonical AGN line).
This result indicates  
that pure LINERs and pure Seyferts are structurally very similar.

\begin{figure}
\centerline{
\epsfxsize=12cm \epsfbox{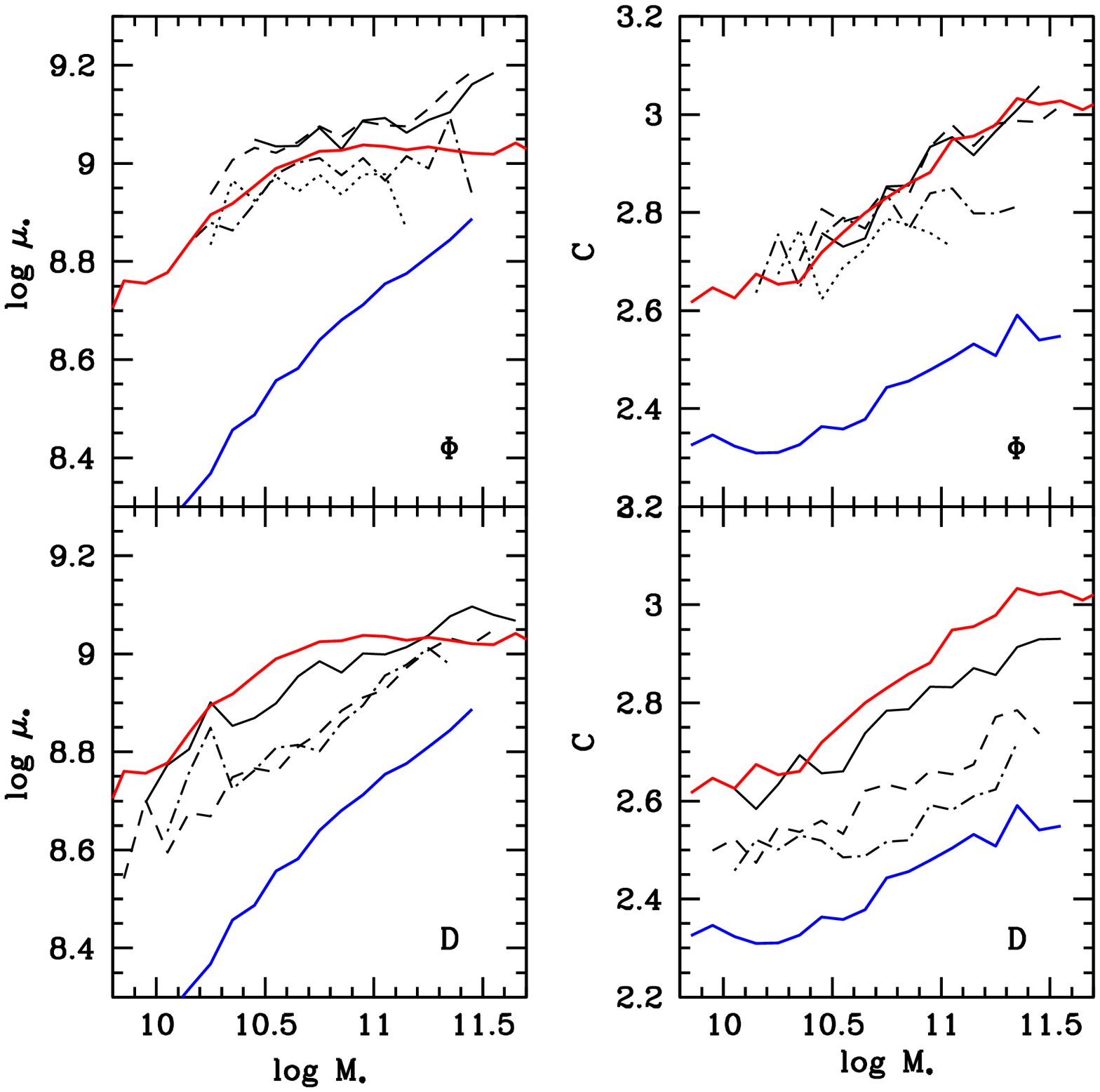}
}
\caption{\label{fig11}
\small
Trends in the $\mu_*-M_*$ and $C-M_*$ relations for AGN hosts as a function of 
 angle parameter $\Phi$ and distance $D$ from the locus of
star-forming galaxies. Blue and red lines show the median relations for
ordinary late- and early-type galaxies. The black line types
are  solid, dashed, dashed-dotted and dotted in order 
of decreasing $\Phi$ (top) and decreasing
$D$ (bottom).}
\end {figure}
\normalsize

\subsection {Mean Stellar Age}

In the previous subsection, we showed that AGN 
occupy host galaxies
with structural properties similar to ordinary early-type galaxies (with
the similarity being stronger for the weak AGN).
The properties of the stellar populations of normal galaxies are known to correlate  
strongly with morphological type. Early-type galaxies have old stellar
populations and very little gas and dust, whereas there is usually plenty of                   
ongoing star formation in late-type galaxies. 
As discussed in Paper I, D$_n$(4000) is an excellent age indicator
at mean  stellar ages  less than $\sim 1$ Gyr (corresponding to D$_n$(4000) values 
$<$ 1.4-1.5). At older ages, the D$_n$(4000) index is sensitive
to metallicity as well as to age. We have shown that AGN are found in massive
($> 10^{10} M_{\odot}$) host galaxies, so metallicity variations among hosts are
likely to be small and it is                                    
appropriate to use the D$_n$(4000) index as a rough stellar age indicator.

In Fig. 12 we plot D$_n$(4000) and H$\delta_A$ as a function of [OIII]
luminosity for all
the AGN in our sample. There is a strong correlation of 
both age-sensitive parameters with AGN luminosity. Only the weakest
AGN have stellar ages in the range that is normal for 
early-type galaxies (D$_n$(4000)$ > 1.7$ and H$\delta_A <$ 1).

\begin{figure}
\centerline{
\epsfxsize=10cm \epsfbox{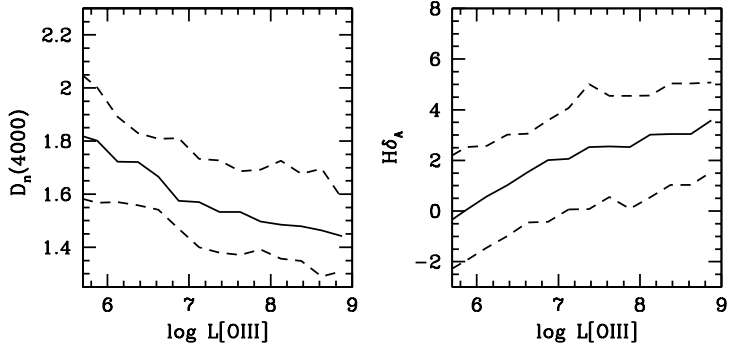}
}
\caption{\label{fig12}
\small
D$_n$(4000) and H$\delta_A$ are  plotted as a function of log L[OIII]. The solid
line shows the median, while the dashed lines indicate the 16-84 percentiles of the
$1/V_{max}$ weighted distribution.}
\end {figure}
\normalsize

In Fig. 13, we compare the  D$_n$(4000)-- 
stellar mass and D$_n$(4000)-- surface density relations for AGN hosts 
with those of normal galaxies. 
The normal galaxy relations exhibit a  strong transition in D$_n$(4000) at 
$M_* \sim 3 \times 10^{10} M_{\odot}$ and  $\mu_* \sim 3 \times 10^8 M_{\odot}$ kpc$^{-2}$.
This reflects a switch from a population of young late-type galaxies with  low masses and
low surface densities            
to a population of old early-type galaxies with  high masses and surface densities.
Fig. 13 shows that strong
AGN do not fit on the relations defined by normal galaxies.
They have a much weaker dependence of age on stellar mass 
or surface mass density than normal galaxies. In particular,
they occur at  much lower values of D$_n$(4000)
than the normal galaxies in the high-mass, high-density regime
($M_* > 3 \times 10^{10} M_{\odot}$, $\mu_* > 3 \times 10^8 M_{\odot}$
kpc$^{-2}$). The weak AGN  resemble normal galaxies more closely.

\begin{figure}
\centerline{
\epsfxsize=15cm \epsfbox{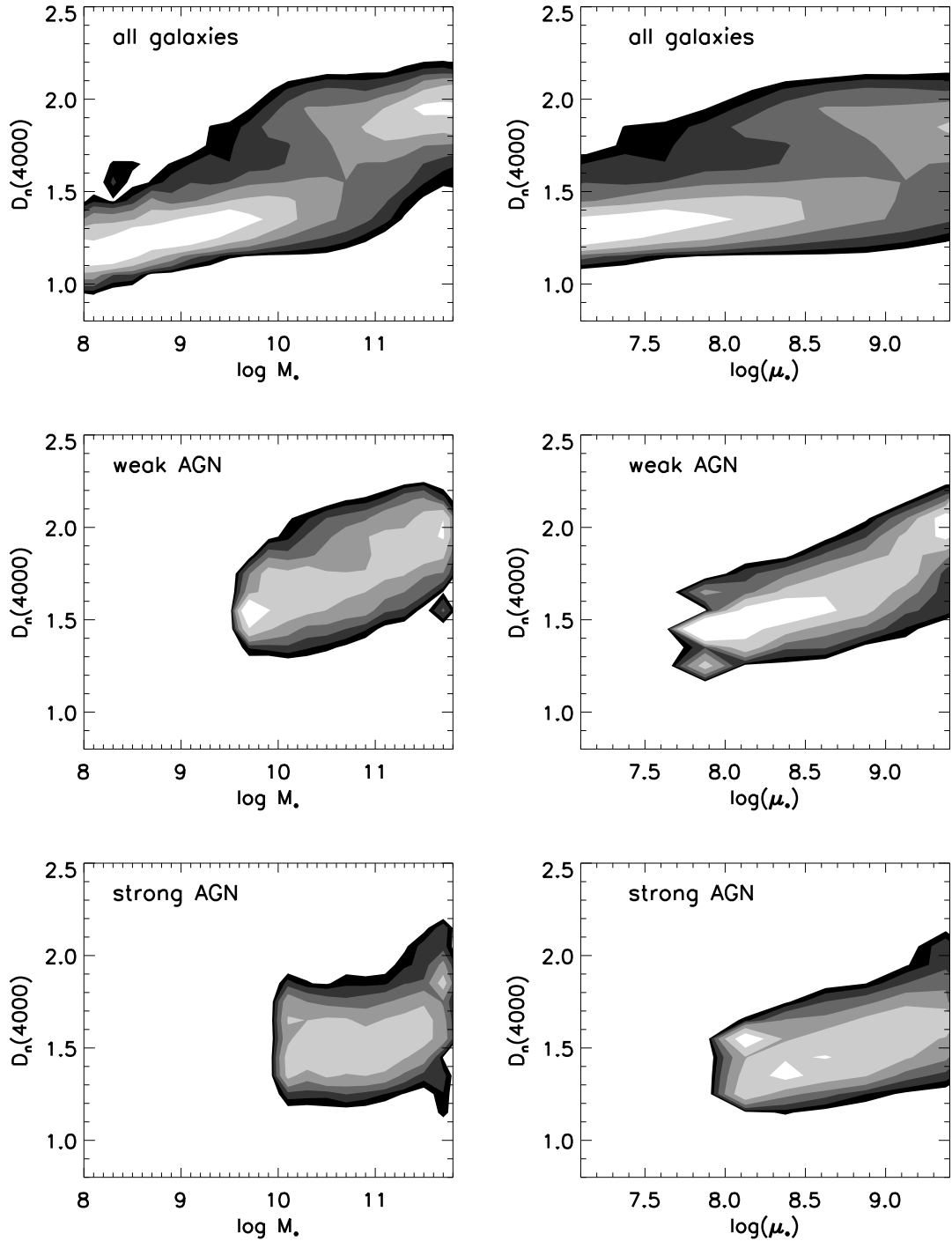}
}
\caption{\label{fig13}
\small
 Conditional density distributions showing trends in D$_n$(4000)  as a function
of stellar mass $M_*$ and surface density $\mu_*$  for ordinary galaxies (top), weak AGN with log L[OIII]$ < 7.0$
(middle) and strong AGN with log L[OIII]$ > 7.0$ (bottom).}
\end {figure}
\normalsize

In Fig. 14, we compare the D$_n$(4000) distributions of AGN hosts with
`normal' early-type ($C>2.6$) and late-type ($C<2.6$) galaxies. This plot is
very similar in spirit to Fig. 10, but reaches a rather different conclusion.       
In Fig. 10, we showed that {\em both}  weak and strong AGN had surface density distributions
very similar to early-type galaxies. Fig.13 shows that
the D$_n$(4000)  distributions of  weak AGN  resemble those of early-types,         
but that strong AGN have stellar ages similar to those of late-type galaxies.

\begin{figure}
\centerline{
\epsfxsize=14cm \epsfbox{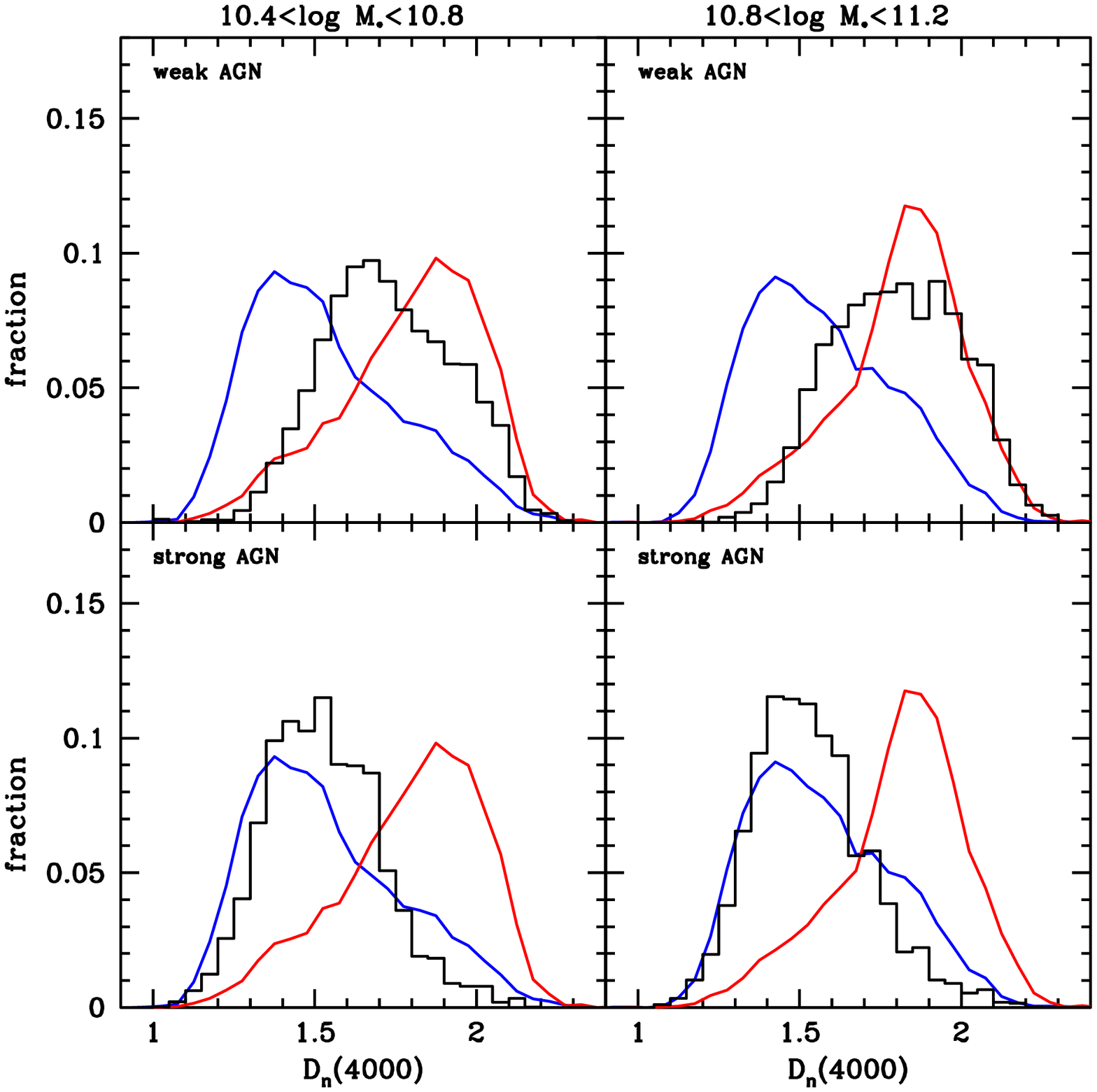}
}
\caption{\label{fig14}
\small
Top: The D$_n$(4000) distribution for the  host galaxies
of weak AGN with log L[OIII]$ < 7.0$  (black) is compared to
those for  early-type ($C>2.6$) galaxies (red) and late-type ($C<2.6$) galaxies (blue)
in 2 separate ranges of $\log M_*$. Bottom: The same, except for
strong AGN with log L[OIII]$ >7.0$}
\end {figure}
\normalsize

As discussed in Paper II, the star formation histories of normal galaxies appear to be
more fundamentally linked to surface mass density than to  mass.
We demonstrated  that there is a ``universal''
relation between D$_n$(4000) and $\mu_*$ that is independent of the stellar mass of the galaxy,
except at the very highest values of $M_*$. The results in Fig. 14 suggest
that the D$_n$(4000)- $\mu_*$ relation for the hosts of strong AGN is significantly different. 
In Fig. 15, we plot the median value of 
D$_n$(4000) as a function of  $\log \mu_*$ for AGN of different luminosities (red lines)
and compare this with the relation obtained for ordinary galaxies (thick black line).
At high surface mass densities, low-luminosity AGN follow the relation
defined by ordinary galaxies, but high-luminosity AGN have substantially smaller
D$_n$(4000) values (a younger age) at a given value of $\mu_*$. At low surface
mass densities, 
the hosts of weak AGN
have older stellar populations than normal galaxies of the same surface density.

\begin{figure}
\centerline{
\epsfxsize=10cm \epsfbox{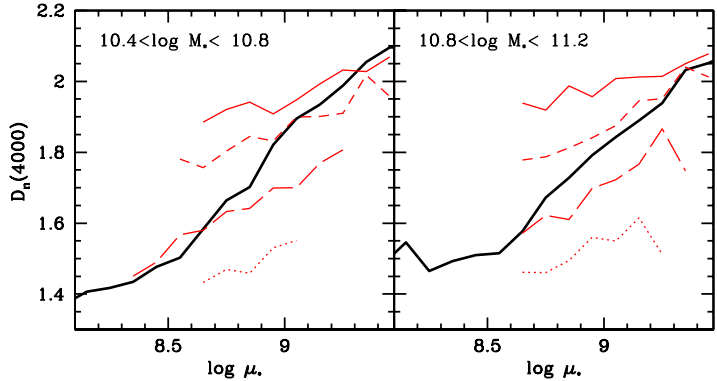}
}
\caption{\label{fig15}
\small
The thick black lines show the median relation between D$_n$(4000) and $\log \mu_*$
for normal galaxies in two different ranges in stellar mass.
The red lines show the median relation for the host galaxies of  AGN with
log L[OIII]$ < 6.0 $ (solid), 6.0$<$ log  L(OIII)$<7.0$ (short-dashed),
7.0$<$ log L[OIII]$<8.0$ (long-dashed) and 
log L[OIII]$>8.0$ (dotted).} 
\end {figure}
\normalsize

How can we reconcile the results of Fig. 15 with the fact that the majority of
galaxies exhibit a very different relation between age and surface mass
density? 
One interpretation is that powerful  AGN reside in early-type galaxies that are
undergoing or have recently undergone a
transient star-forming event (starburst). The magnitude of this event
might plausibly scale with AGN luminosity, so that its effects on
the stellar population and structural parameters are most pronounced
in the hosts of strong AGN. The duration of this event is short
compared to the age of the Universe, so only a fraction of galaxies are
caught in the act. As the AGN switches off and the stellar population ages,
the galaxy eventually
(re)joins the ranks of the normal early-type population.
If this hypothesis is correct, then we ought to see
evidence that the
star formation timescales in high luminosity AGN are short.

In Fig.16, we show the relation between D$_n$(4000) and the position of the galaxy in the
BPT diagram. Because D$_n$(4000) is itself very strongly correlated with [OIII] luminosity 
(Fig. 12), 
we remove this effect by analyzing weak AGN and strong AGN separately.
The left panel of Figure 16. shows that there is then no residual correlation of
D$_n$(4000) with $\Phi$.
The  correlation of  D$_n$(4000) with distance parameter  $D$ is more interesting.                              
The right panel of Fig. 16 shows that in weak AGN,  D$_n$(4000) 
correlates strongly  with  $D$ , but in strong AGN, the dependence is much                 
weaker. 
So far we have interpreted  $D$  as a measure of the 
relative contribution of star formation to the observed emission
lines.  (Recall that we showed in Fig. 3 that AGN luminosity does not correlate with $D$,
implying that galaxies with lower values of $D$ have relatively  more star formation).
For weak AGN, the increase in D$_n$(4000) with distance from the
locus of star-forming galaxies appears to support this hypothesis,
but how can we explain the fact that there is a much weaker  trend for strong AGN?
The parameter  $D$ is determined by the amount of
on-going star-formation in the galaxy because the Balmer emission-lines are excited
by O stars with lifetimes of several million years.  D$_n$(4000), on the other hand,
tracks the luminosity-weighted mean stellar age on much longer timescales
($\geq 10^8$ years). Thus, the result in Fig. 16 can be understood if
high-luminosity AGN with young stellar populations (small D$_n$(4000))
and large $D$ have recently stopped forming stars
(i.e. they are {\em post-starburst}
systems). In the next section, we demonstrate that this is indeed the case.

\begin{figure}
\centerline{
\epsfxsize=12cm \epsfbox{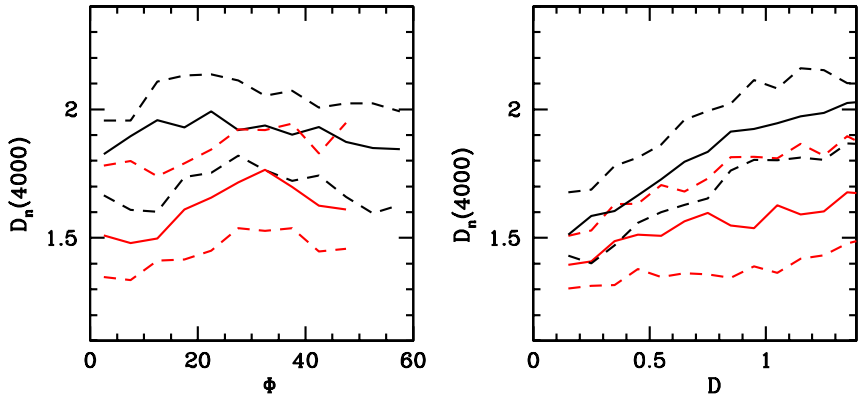}
}
\caption{\label{fig16}
\small
D$_n$(4000) is plotted as a function of $\Phi$ (for AGN lying above the
Kewley et al line) and $D$ for weak AGN (black) and for
strong AGN (red). The solid lines indicate the median and the dashed lines
the 16-84 percentiles of the $1/V_{max}$ weighted distribution.}
\end {figure}
\normalsize

\subsection {The Star Formation Histories of AGN}

Strong H$\delta$ absorption arises in galaxies that experienced a burst of star formation that ended
$0.1-1$ Gyr ago. As discussed in Papers I and II, the location of galaxies in the
D$_n$(4000)/H$\delta_A$ plane is a powerful diagnostic of whether they have been forming
stars continuously or in bursts over the past 1-2 Gyr. Galaxies with continuous star formation
histories occupy a very narrow strip in this plane.  A recent burst is {\em required} in order
to obtain  significant displacement away from this locus. 
In Paper II, we also showed that the
probability for a galaxy to have experienced  a recent burst is a
strong function of stellar mass. A  significant
fraction (more than 10\%) of low-mass galaxies with
$M_* \sim 10^8 M_{\odot}$   appear to have experienced recent bursts. 
The fraction of bursty galaxies drops by more than a factor of 10 for galaxies more massive than                 
$10^{10} M_{\odot}$.
In the previous section, we showed that the host galaxies of powerful AGN  had mean
stellar ages that were similar to those of ordinary late-type galaxies.
In this section, we ask whether the star formation
histories of AGN hosts are different from those of              
ordinary galaxies.  In particular, we ask whether there is any evidence that powerful AGN are associated
with bursty rather than continuous star formation histories.

In the SDSS sample, the typical error on the H$\delta_A$ index  is 1.4 \AA, which is large
compared to the total range of values spanned by this index (-3 to +9 \AA).
In order to analyze the star formation
histories of AGN in a more reliable way, we have extracted a subsample of 
objects  with small ($< 0.5$ \AA) errors in 
H$\delta_A$. This reduces our total sample of AGN  to 1400 objects.  
Compared to the full sample of 22,000
AGN, our subsample is biased to lower mean redshift (0.05 rather than 0.1).
Thus, the spectra
sample systematically smaller regions of the host galaxies than in the full
sample (typical projected fiber diameter $\sim$3 kpc {\it vs.} $\sim$6 kpc).
However, so long as we impose the same
selection criterion on the  normal galaxies as we do on the AGN, the comparison between
the two subsamples is a fair one.               

In Fig. 17, we plot our subsamples of normal galaxies, strong AGN and weak AGN in the
D$_n$(4000)/H$\delta_A$ plane. The locus occupied by galaxies with
continuous star formation histories is indicated using red coloured symbols.
The normal galaxies are selected to have stellar masses in the 
same range as the AGN hosts
($10^{10}-5 \times 10^{11} M_{\odot}$). As can be seen, most normal galaxies
fall close to the locus of continuous star formation histories.
A significant fraction of the powerful AGN, on the other hand, are displaced to
high values of H$\delta_A$. This is shown quantitatively in Fig. 18
where we plot the fraction $F$  of AGN with H$\delta_A$ values that are displaced
by more than 3$\sigma$ above the locus of continuous models.         
This fraction increases from several percent for the weakest AGN to nearly
a quarter for the strongest.
The corresponding value of $F$ for the sample of     
normal massive galaxies is only 0.07. For normal massive galaxies with
`young' stellar populations (D$_n$(4000)$<1.6$), $F=0.12$.

\begin{figure}
\centerline{
\epsfxsize=14cm \epsfbox{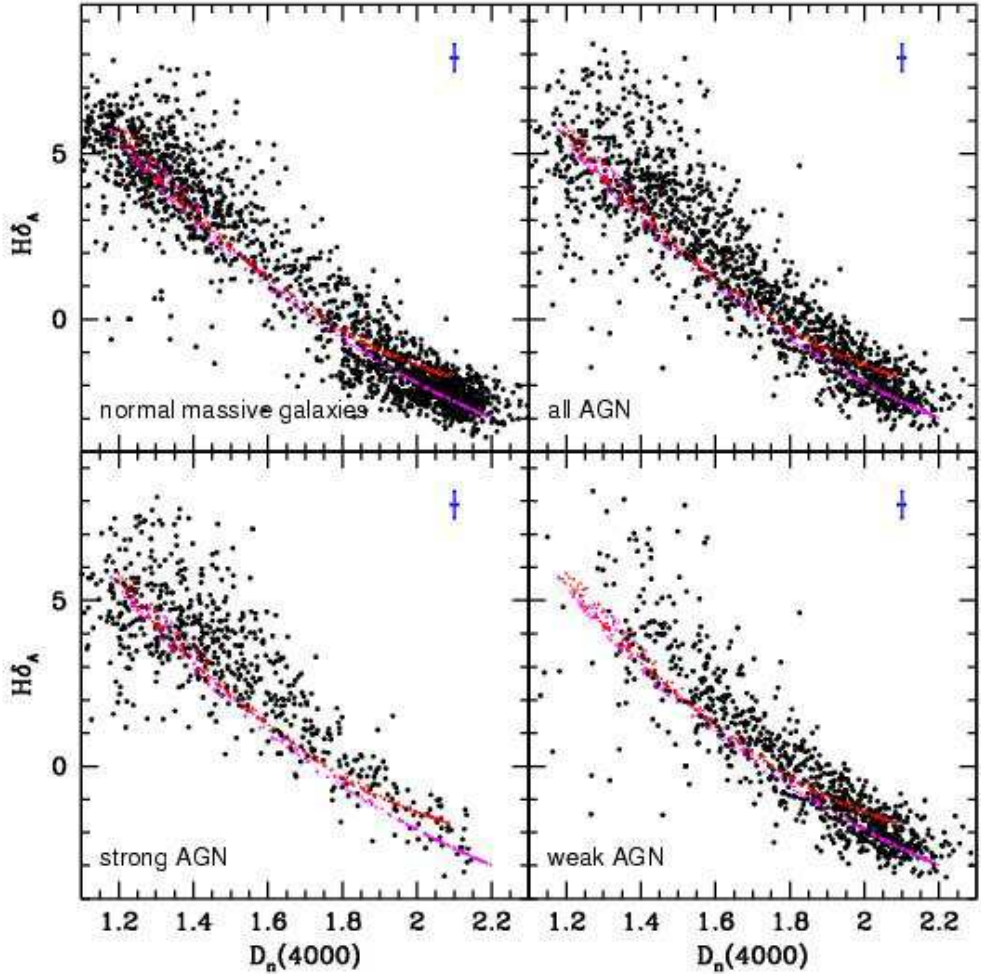}
}
\caption{\label{fig17}
\small
High $S/N$  subsamples of normal galaxies, strong AGN and weak AGN are plotted in the
D$_n$(4000)/H$\delta_A$ plane. The locus occupied by galaxies with
continuous star formation histories is indicated using  coloured symbols.
Red symbols are for solar metallicity models and magenta symbols indicate models with
twice solar metallicity.
The average errorbar on the indices is shown in the corner of each panel.}
\end {figure}
\normalsize

\begin{figure}
\centerline{
\epsfxsize=7cm \epsfbox{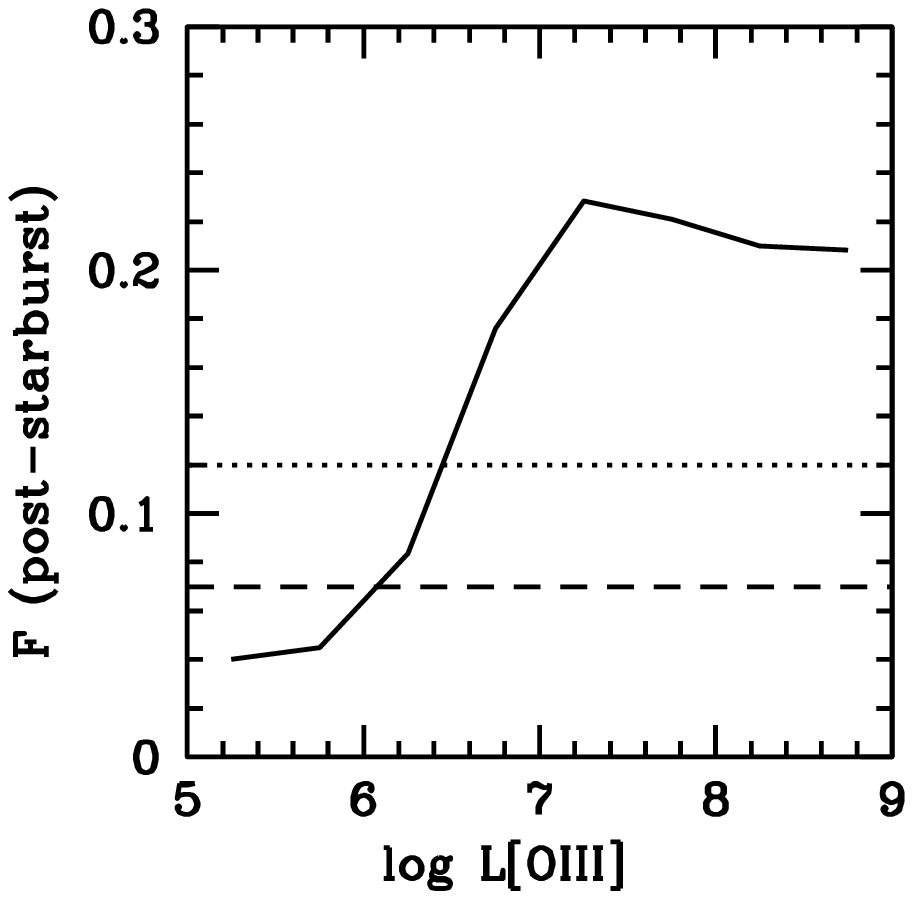}
}
\caption{\label{fig18}
\small
The fraction $F$  of AGN with H$\delta_A$ values that are displaced
by more than 3$\sigma$ above the locus of star-forming galaxies is
plotted as a function of log L[OIII].
The dashed line indicates the fraction of such systems in the
subsample of normal massive galaxies. The dotted line
indicates the fraction of such systems in the subsample
of normal massive galaxies with D$_n$(4000)$< 1.6$.}
\end {figure}
\normalsize

In Fig. 19, we colour-code the AGN according to $D$, the distance from the
locus of star-forming galaxies in the BPT diagram. AGN that lie close to the star-forming
locus are coloured  blue, while those that lie far away are coloured red.
Nearly all the weak AGN with large values of $D$ have
large values of D$_n$(4000). Most of the strong AGN with large values of $D$
have higher-than-normal values of
H$\delta_A$.  
This lends further credence to our hypothesis that the  latter objects are
post-starburst systems; they experienced a burst some time
in the past and are now evolving towards the locus occupied
by ordinary early-type galaxies.

\begin{figure}
\centerline{
\epsfxsize=10cm \epsfbox{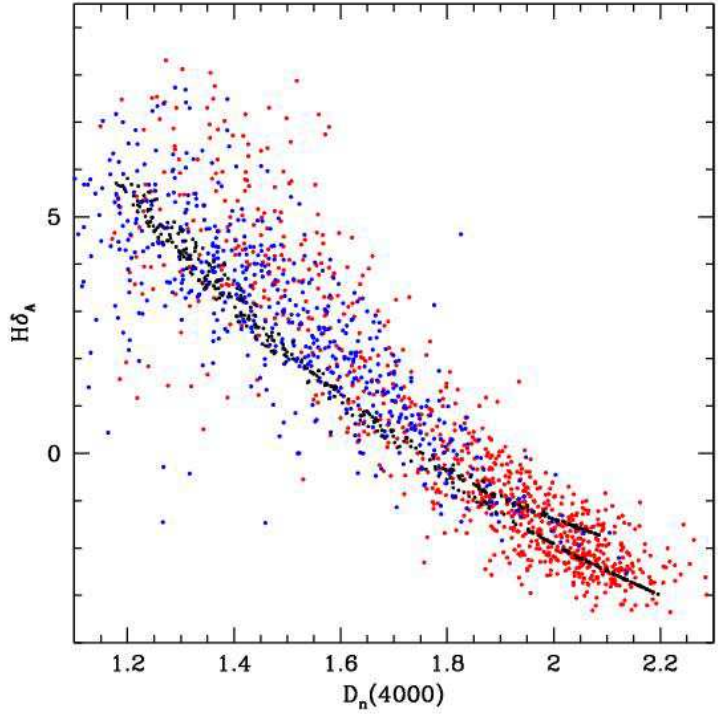}
}
\caption{\label{fig19}
\small
AGN are plotted in the D$_n$(4000)/H$\delta_A$ plane and are colour coded according to the value of
$D$, the distance from the locus of star-forming galaxies.
Blue is for $D< 0.6$ and red is for $D>0.6$.}
\end {figure}
\normalsize

\subsection {Radial Distribution of the Star Formation}

Most previous studies of the stellar populations of type 2 AGN hosts have focused on the
{\em nuclear regions} of the host galaxies (see Heckman 2003 for a recent
review). The stellar populations in the nuclei of weak AGN are
predominantly old (e.g. Ho, Filippenko, \& Sargent 2003), while
a young stellar population is clearly present in about half of
powerful type 2 Seyfert nuclei (e.g. Cid Fernandes et al. 2001;
Joguet et al. 2001).
We obtain qualitatively similar trends in mean stellar age as a function of 
AGN luminosity, but
we find that there are young stars in most  AGN
with [OIII] luminosities greater than $10^7 L_{\odot}$.
One might thus speculate that there might be  systematic radial
gradients in the stellar populations of the host galaxies of these systems.

We can test this by splitting our sample into different bins in normalized
distance $z/z_{max}$.
We then find that there are
rather strong radial gradients in mean stellar age and that the youngest stars
appear to be located  well {\em outside} the  nuclei of the host galaxies.
This is illustrated in Fig. 20. We have selected AGN with [OIII] luminosities in excess
of $3 \times 10^7$ $L_{\odot}$ and we plot the locations of the host
galaxies in the $C$-D$_n$(4000) plane. The left panel shows objects with
$z/z_{max}>0.8$. These galaxies have  redshifts in the range $0.1-0.2$ 
and the spectra sample between 40-60\% of the total light. As can be seen,
the vast majority of these objects have D$_n$(4000) values less than 1.6
and are hence younger than the typical early-type galaxy in our sample.
The right panel shows objects with $z/z_{max}<0.3$. These galaxies have typical 
redshifts of $0.03-0.04$ and the spectra sample the light from the inner
1-2 kpc of the host galaxy. As can be seen, more than half the galaxies
now have D$_n$(4000)$>1.6$. The shift in mean stellar age occurs both in
late-type ($C<2.6$) and in  early-type ($C>2.6$) AGN hosts.
We conclude the star formation associated with AGN activity
is not primarily confined to the nuclear regions of the host galaxy. 

\begin{figure}
\centerline{
\epsfxsize=10cm \epsfbox{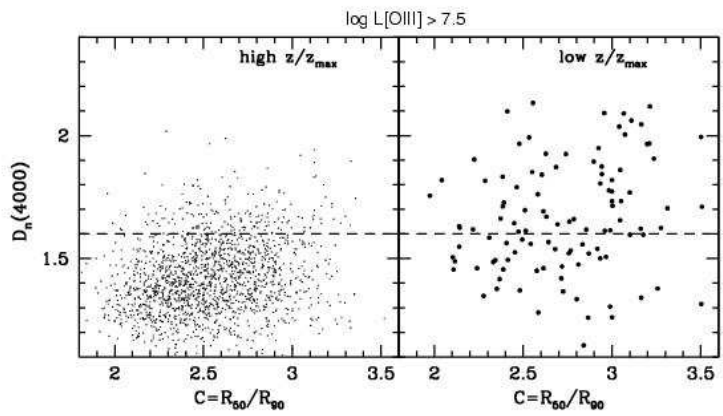}
}
\caption{\label{fig20}
\small
AGN with [OIII] luminosities greater than
$10^{7.5}$ L$_{\odot}$  are plotted in the  D$_n$(4000)/$C$ plane.
In the left panel we show objects with $z/z_{max}>0.8$. In the right
panel, we show objects with $z/z_{max}<0.3$.} 
\end {figure}
\normalsize

\subsection {Dust}
As discussed in section 3, we correct the [OIII] luminosities of the AGN in our sample
for dust using  the difference between the observed
H$\alpha$/H$\beta$ emission line flux ratios and the case-B recombination value
(2.86). We assumed  an attenuation law of the
form $\tau_{\lambda} \propto \lambda^{-0.7}$ (Charlot \& Fall 2000). 
In the top right panel of Fig. 21, we plot the average attenuation  at [OIII]$\lambda$5007
as a function of normalized AGN luminosity. As can be seen, the correction is
very small for weak AGN, but rises strongly with increasing L[OIII],  and reaches
2.5-3 magnitudes for the most powerful AGN. This is as expected, since
weak AGN have old stellar populations (thus, little dust), whereas
strong AGN have young stellar populations (more dust).
Is 3 magnitudes of extinction at [OIII] a reasonable value for the 
strong AGN? 
Dahari \& De Robertis (1988) have  tabulated the Balmer decrements
of 64 type 2 Seyfert nuclei, and the implied mean [OIII] attenuation
is $\sim$2.5 magnitudes. This is  similar to what we measure
through much larger physical apertures.
In the left panel of Fig. 21, we plot the 
attenuation as a function of stellar
mass for normal  emission-line galaxies in our sample.
The average value of the attenuation at 5007 \AA \hspace{0.1cm} 
for massive ($>$ few $\times 10^{10} M_{\odot}$) star-forming  galaxies is
the same as that for the powerful AGN in our sample. These
comparisons give us confidence
that our average dust corrections are appropriate.

In the bottom panels of Fig. 21 , we plot the attenuation in AGN hosts as a function of 
$D_n(4000)$ and $D$. There is a clear connection between star formation
and extinction in AGN hosts, with younger (low D$_n$(4000)) and
more strongly star-forming (low $D$) galaxies exhibiting more attenuation due
to dust.

\begin{figure}
\centerline{
\epsfxsize=11cm \epsfbox{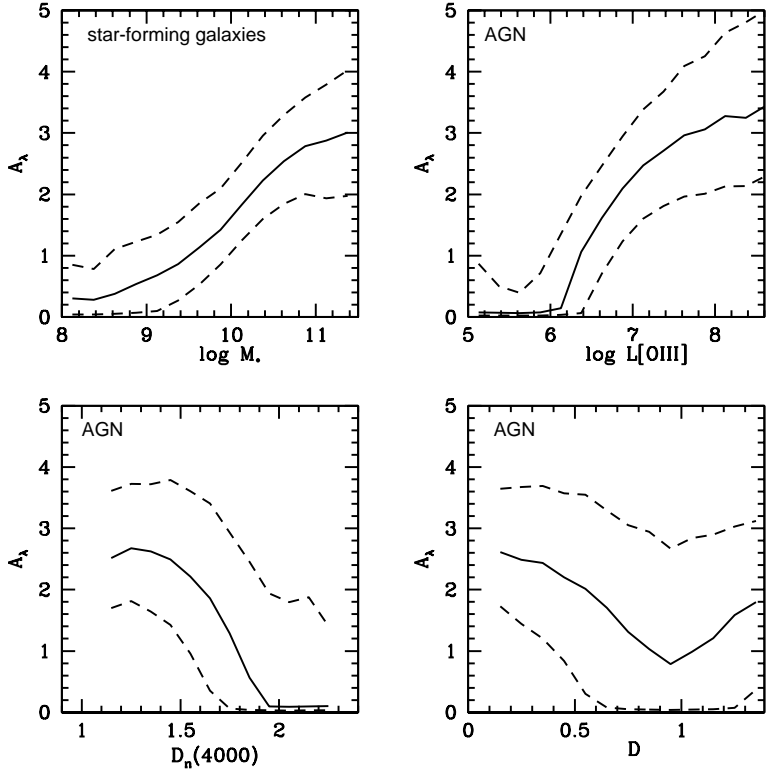}
}
\caption{\label{fig21}
\small
{\bf Top Left}: The attenuation  at [OIII]$\lambda$5007 in magnitudes
is plotted as a function of stellar mass for normal star-forming galaxies.
{\bf Top Right}: The attenuation is plotted as a function of  [OIII]   
luminosity for AGN. {\bf Bottom}: Attenuation in AGN hosts is plotted as a function
of D$_n$(4000) (left) and distance parameter $D$ (right). 
The solid line shows the median value of the attenuation , while
the dashed lines show the 16-84  percentiles of the
$1/V_{max}$ weighted distribution.} 
\end {figure}
\normalsize

\section{Comparison of Type 2 AGN and  QSO Hosts}

We have shown that powerful type 2 AGN reside in
massive galaxies with young stellar populations. So far, we have
not considered type 1 Seyfert nuclei and QSOs because their
optical continuum is primarily produced by the AGN (rather than
by stars).

In the simplest version of the ``unified'' model
(Antonucci 1993), type 1 and type 2 AGN are drawn from
the same parent population and differ only in our viewing angle
with respect to a circumnuclear obscuring medium.
In this case, the stellar contents of the two types should be the same.
On the other hand, if the solid angle covered by the obscuring
medium varies substantially from one system to another, then 
the type 2 AGN would be preferentially drawn from the more obscured
systems. In this case,
there could be systematic differences in  stellar content 
between type 1 and type 2 AGN, since 
the covering factor of the dusty material might be related to the amount
of star formation in the galaxy.
There have been suggestions that this effect is present in  Seyfert galaxies
(e.g. Malkan, Gorjian, \& Tam 1998; Maiolino et al. 1995; Oliva et al. 1999).
Dunlop et al. (2001) have argued that powerful
low-redshift QSOs are hosted by relatively normal old massive elliptical
galaxies. This would make them fundamentally different from the hosts
of powerful type 2 AGN.

In this section, we compare the stellar populations of the hosts 
of powerful type 1 and type 2 AGN using
our present sample and a comparison
sample drawn from the SDSS QSO spectroscopic
survey ( Schneider et al, in preparation; see Richards et al. (2002) for details on how QSOs
are selected). The stellar features in QSOs are diluted
by the addition of the QSO continuum, and measurements of them are
noisier. We therefore 
combined several hundred individual spectra in order to increase the
signal-to-noise. Specifically,
we selected all QSOs
having $z \leq$ 0.2 and $L[OIII] \geq 10^{41}$ erg~s$^{-1}$.
These selection criteria are designed to maximize the overlap with the
type 2 AGN population our sample.
For each QSO, we find a type 2 AGN with a similar redshift
and [OIII] luminosity.
The sample comprises 449 QSOs and 449 type
2 AGN. The median [OIII] luminosity is $10^{41.3}$ erg s$^{-1}$.
\footnote{
It is very difficult to correct the [OIII] luminosities of QSOs for
extinction, because the Balmer lines are dominated
by emission from the Broad Line Region.
We have therefore chosen to match the type 2 AGN and QSOs in
terms of their observed [OIII] line luminosities. We thus
neglect any differences in extinction  between the two
kinds of objects. For reference, the median extinction-corrected
[OIII] luminosity for the 449 type 2 AGN is $10^{42.1}$ erg s$^{-1}$
= $10^{8.5} L_{\odot}$. These are among the most powerful AGN in our sample.}
The median absolute magnitude of the QSO sample is $M_r$ = -21.6, or
about one magnitude fainter than the corresponding value for the
radio-quiet QSOs studied by Dunlop et al. (2001).

The individual flux-calibrated spectra were normalized at
5500 \AA\ and combined with no additional weighting.
The results are shown
in black (QSOs) and in red (matched type 2 AGNs) in the top panel
of Fig. 22. 
For comparison, we plot in blue the composite spectrum of 60 of the
quasars in our sample with the highest
{\em continuum} luminosities at 5007 \AA\ and
with $0.1 < z < 0.35$. We will refer to this composite as
the ``pure QSO'' spectrum.

As can be seen, the
stellar absorption features in the matched QSO spectrum are weaker
than in the type 2 AGN spectrum. This is expected, because a substantial
fraction of the continuum in the matched QSO spectrum is coming from
a central nuclear source rather than from stars.
This is illustrated in detail in the central
panel of Fig. 22, where we plot
in green the linear combination of the type 2 AGN + pure QSO spectra
that produces the best fit to the matched QSO spectrum.
We find  a very good overall fit if 30\% of the continuum
at 5500 \AA\ comes from a pure QSO. The main residuals are located around
the broad emission line features, which show substantial variations from
one QSO to another.
In the bottom panel of   Fig. 22, we  show a blow-up of the region of
the spectrum between 3700 and 4000 \AA.  We have labelled  the main
stellar absorption line features (these include a series of
high-order Balmer lines and the Calcium H+K feature).  It is clear that
the stellar absorption spectra are virtually identical in type 2 AGNs and
in the matched QSOs.

We therefore conclude that there is no
significant difference in stellar content
between the host galaxies of type 2 AGNs
and QSOs with the same [OIII] luminosity and redshift. This is consistent
with the predictions of the unified model for AGN, and establishes
that  a young stellar population
is associated with all types of  AGN with strong [OIII] emission.

\begin{figure}
\centerline{
\epsfysize=16cm \epsfxsize=13cm  \epsfbox{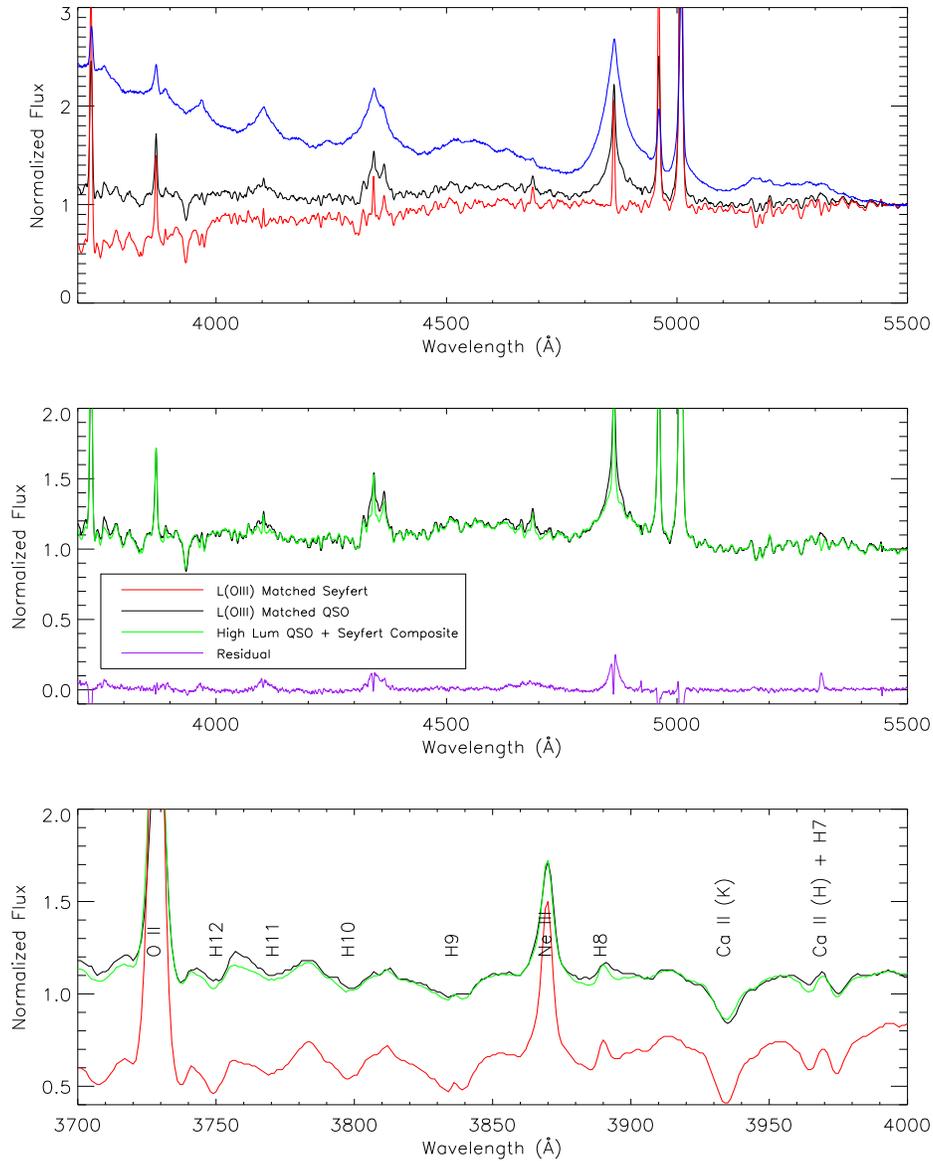}
}
\caption{\label{fig22}
\small
{\bf Top:} The composite type 2 AGN spectrum is shown in red. The composite spectrum
of matched
QSOs is shown in black. A composite spectrum of  QSOs with very high continuum
luminosities is shown in blue.
{\bf Middle:} A linear combination of   70 \% type 2 AGN + 30 \% pure QSO (green ) is
compared to the matched QSO spectrum (black).
The residual is shown in purple.
{\bf Bottom:} The spectral region between 3700 and 4000 \AA\ is shown in more detail.
The main stellar aborption lines, including
the higher-order Balmer lines and Calcium H+K lines  have been marked
for clarity.}
\end {figure}
\normalsize

\section {Discussion}

Our main results on the properties of AGN host galaxies can be summarized as follows:
\begin {itemize} 
\item {\bf Stellar Masses.} Type 2 AGN reside almost exclusively in massive
galaxies.  The fraction of galaxies with AGN    
declines very strongly at stellar masses below  $10^{10} M_{\odot}$.

\item {\bf Structural Properties.} Type 2 AGN occupy host galaxies with very     
similar sizes and stellar surface mass densities to  normal ``early-type'' galaxies
in our sample. The structural properties of AGN hosts depend very little on AGN power.

\item {\bf Mean Stellar Ages.}The age distribution of type 2  AGN hosts
is a strong function of [OIII] luminosity. Low luminosity AGN have old stellar populations
similar to those of  early-type galaxies. High luminosity  AGN reside in significantly
younger hosts and have  D$_n$(4000) distributions
similar to those of  normal late-type ($C<2.6$) galaxies.     

\item {\bf Starbursts.} Although powerful type 2 AGN have stellar ages similar to
massive late-type galaxies,  a  larger fraction appear to  have experienced
significant bursts of star formation in the past one or two Gyr.    

\item {\bf Star Formation Outside the Nucleus} The star formation in powerful AGN is not           
concentrated primarily in the nuclear regions of the galaxy, but is distributed over
scales of at least several kpc.

\item {\bf Ongoing Star Formation.} We interpret the parameter $D$ (defined as
the distance from the locus of star-forming galaxies in the BPT diagram)  as a qualitative 
measure of the relative  amount of ongoing star formation in the galaxy. We find that type 2 AGN with
small values of $D$  have lower stellar surface mass densities
and concentrations. As expected, they also have more dust and younger mean stellar ages.

\item {\bf Ionization State.} We use a position angle $\Phi$ to characterize the ionization state of
the AGN in our sample. We find that $\Phi$ is strongly correlated with AGN luminosity.
Low-luminosity AGN have low-ionization ``LINER-type'' emission line ratios.
High-luminosity AGN have high-ionization ``Seyfert-type'' line ratios. There is a
sharp discontinuity in the  [OIII] luminosity distribution at $\Phi \sim 25^{\circ}$,
suggesting that AGN split into two distinct populations in the BPT diagram.

\item {\bf Young stars in type 1 AGN.} 
There is no significant difference in stellar content 
between the host galaxies of strong type 2 AGN     
and QSOs with the same [OIII] luminosity and redshift. 
\end {itemize}

Our most striking result is that powerful AGN are found in ``young bulges'', i.e.
massive galaxies with stellar surface densities similar to early-type galaxies, but
with young stellar populations.
Ordinary galaxies with these characteristics are rare in the local Universe. We illustrate this
graphically in  Fig. 23.
The top panels show how the the total stellar  mass density in galaxies is
partitioned among galaxies as a function of $M_*$, $\mu_*$ and D$_n$(4000). The 
middle panels show how the
total [OIII] luminosity density contributed by AGN  is distributed. 
As can be seen, most of the [OIII] luminosity
density comes from galaxies with $M_*$ in the range
$3 \times 10^{10}-10^{11} M_{\odot}$,   $\mu_* \sim 10^9 M_{\odot}$ kpc$^{-2}$,
and D$_n$(4000)$<1.6$.
However, there is rather little  stellar mass located in this region of parameter space.             
At the current epoch, most of the stars located in massive, high surface density galaxies are old.
The bottom panels show how the fraction of galaxies that host
 AGN with L[OIII]$> 10^7 L_{\odot}$ varies as a function
of $M_*$, $\mu_*$ and D$_n$(4000). In Fig. 5 we showed that
$F_{AGN}$ defined in this way did not depend on redshift (and hence aperture).  
For  galaxies with $M_* \sim 10^{11} M_{\odot}$, $F_{AGN} \sim 0.1$.
Fig. 23 shows that for galaxies in this
mass range with a young stellar population ($D_n(4000) < 1.5$), 
the AGN fraction reaches values in excess of       
a third.

\begin{figure}
\centerline{
\epsfxsize=15cm \epsfbox{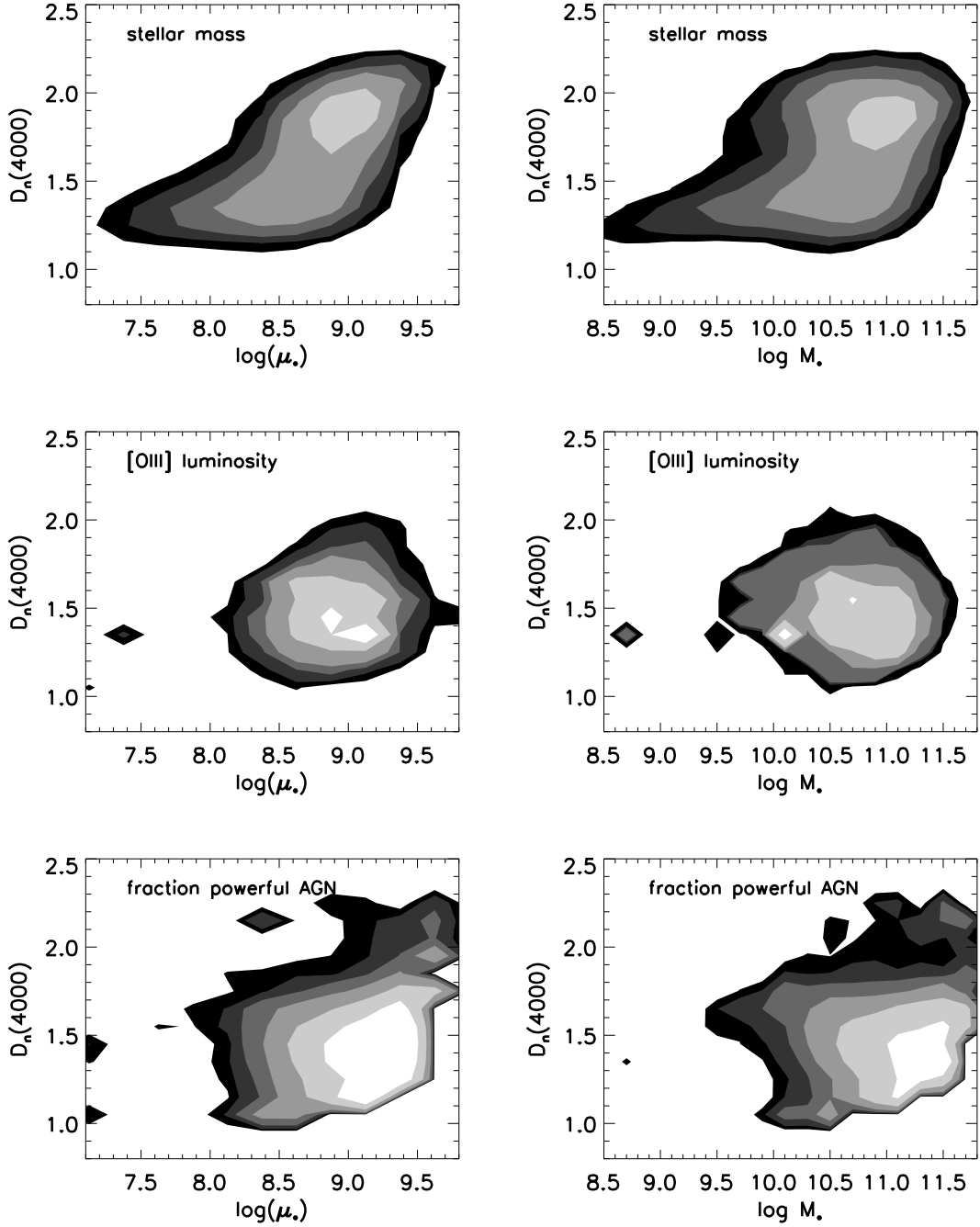}
}
\caption{\label{fig23}
\small
{\bf Top:} The distribution of the total stellar  mass density in the 
D$_n$(4000)/ $\mu_*$ and D$_n$(4000)/$M_*$  planes. 
{\bf Middle:} The distribution of the total [OIII] luminosity density in AGN
in the same planes. Contours indicate a factor two decrease in
in density. {\bf Bottom:} The fraction of galaxies that host
powerful (L[OIII]$> 10^7 L_{\odot}$) AGN. The contours represent AGN fractions of
0.3, 0.15, 0.07, 0.035, 0.017 and  0.008.}
\end {figure}
\normalsize

One interesting question is how our results fit in with other studies of active galaxies.
It has often been speculated that nuclear activity  and strong starbursts
are both triggered by mergers or interactions between galaxies.
Genzel et al (2001) and Tacconi et al (2002) have recently carried out a near-infrared spectroscopic study
of a sample of ultraluminous infrared galaxies. They find that the ULIRG velocity dispersion distribution and
their location on the fundamental plane  closely resemble those of intermediate mass elliptical galaxies.
The star formation rates in ULIRGS are typically around a  hundred  solar masses per year. 
A significant fraction of these objects host 
active nuclei and many are observed to be part of interacting or merging systems (e.g. Bushouse et al 2002).

Canalizo \& Stockton (2001) have recently studied a sample of optically-selected QSOs
that have far-infrared colours  similar to those of  ULIRGs.                                                
They propose that these objects are {\em transition systems.}                                      
The scenario they explore is  one proposed
by Sanders et al (1988), where  ULIRGS and QSOs represent different phases of a single
type of event. In the Sanders et al picture,  the merging of two  galaxies causes gas to 
flow  towards the nuclear regions, triggering both starburst
activity and the accretion of gas onto the  black hole.  Initially, 
the nuclear region of the merger remnant is shrouded in dust and the AGN is hidden from view. 
Later, as the galaxy sheds its dust cocoon, ionizing photons are able to
escape, 
and extended [OIII] emission regions appear. The end-product of the merging   
event is a massive early-type system.  

Most  imaging studies of nearby type 2 AGN have shown that unlike 
the ULIRGs, only a minority  of these objects
exhibit obvious signs of an ongoing  interaction or merger 
(e.g. de Robertis, Hayhoe \& Yee 1998 and references therein). Instead, they are typically
early-type galaxies with relatively normal morphologies.
It is important to note that the space density
of the AGN in our sample is a factor 100-1000 times greater than that of
ULIRGs or of bright QSOs and it is therefore 
likely  that only a restricted subset of these objects are analogues of bright QSOs or ULIRGs.
A major merger is not the only process that could bring fuel to a central black hole.
Minor mergers and bar-induced inflows are frequently cited
mechanisms that
might funnel some
gas to the nucleus, thereby temporarily activating an otherwise normal galaxy,
and at the same time triggering star formation that could contribute
to building the bulge.  

The connection between star formation and the AGN is clearest for the most powerful AGN in our sample
 ($L[OIII] \geq 10^{7} L_{\odot}$). We have found that the hosts
of these strong AGN are either galaxies with considerable on-going 
star-formation
or are post-starburst systems with ages of $\sim10^8$ to
$10^9$ years.  We have extracted SDSS images of $\sim$100 nearby 
($z <$ 0.1) AGN with
L[OIII]$> 10^{7.8} L_{\odot}$  and we find that the host galaxies
 fall into three broad categories:
\begin {enumerate}
\item {\em Single blue spheroidal/amorphous galaxies:}
These are symmetric with no close companions and no
significant structure. They look like normal elliptical/S0 galaxies except that 
they have anomalously
blue colours. In total they make up around 40\% of the sample.
\item {\em Single disk galaxies:} These differ from the above in that structure outside the
  bright core is visible (e.g. arms, bars, dust lanes). No close companions
  are visible.  These systems make up $\sim$ 30\%
  of the sample.
\item  {\em Disturbed/interacting Galaxies:} These have one or more
close companions, and/or show obvious tidal debris. They make up
$\sim 30$\% of the sample. 
\end {enumerate}
In Figs. 24 and 25 we present a montage of the basic types of system found in our
sample of
powerful AGN.

\begin{figure}
\centerline{
\epsfysize=18cm \epsfxsize=12cm  \epsfbox{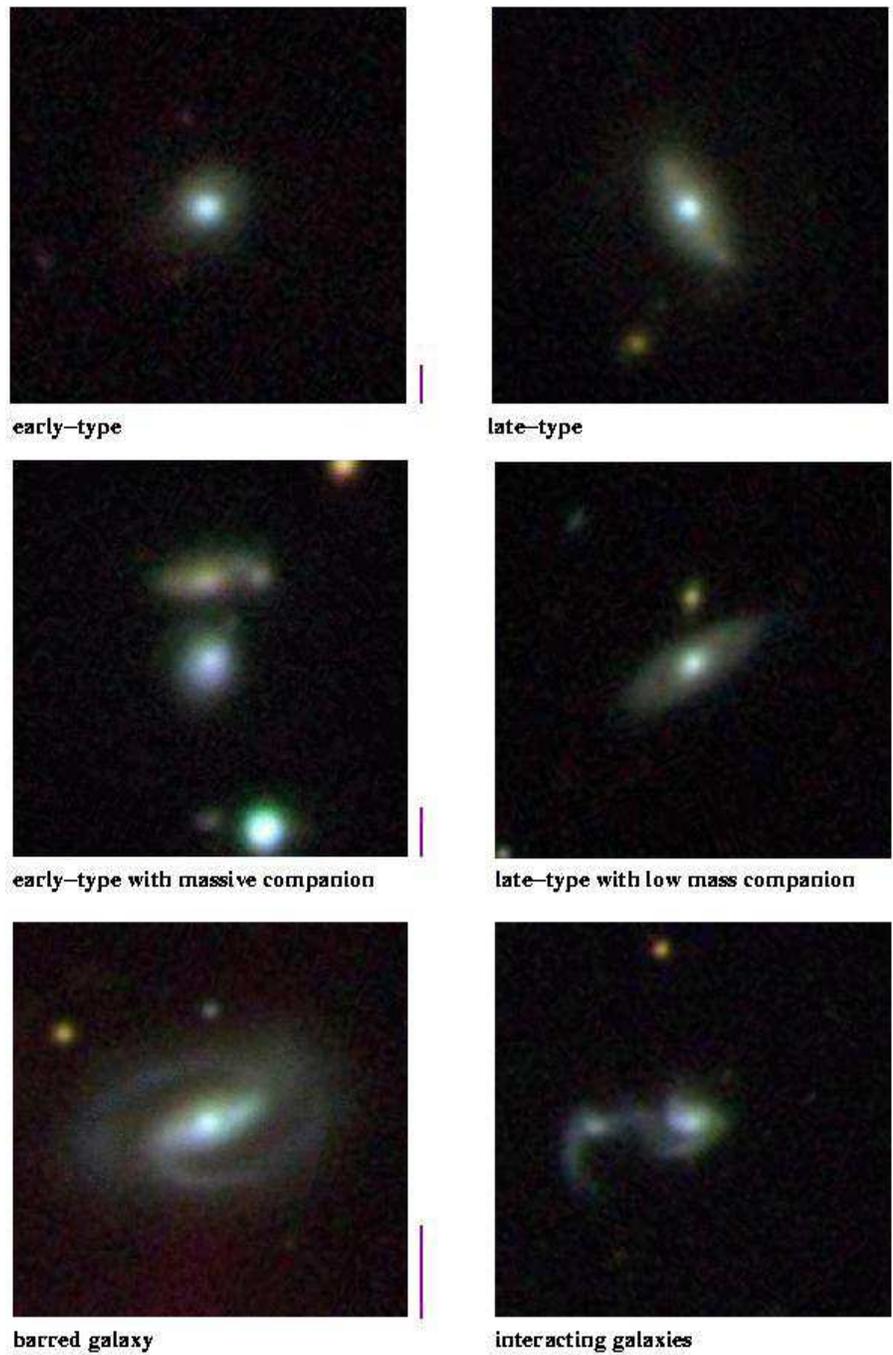}
}
\caption{\label{fig24}
\small
A montage of 48"$\times$48"  images of AGN with 
[OIII] luminosities greater than $10^{7.8} L_{\odot}$. The purple
bar next to each image is 10 kpc long  at the redshift of the AGN.}
\end {figure}
\normalsize

\begin{figure}
\centerline{
\epsfysize=12cm \epsfxsize=12cm  \epsfbox{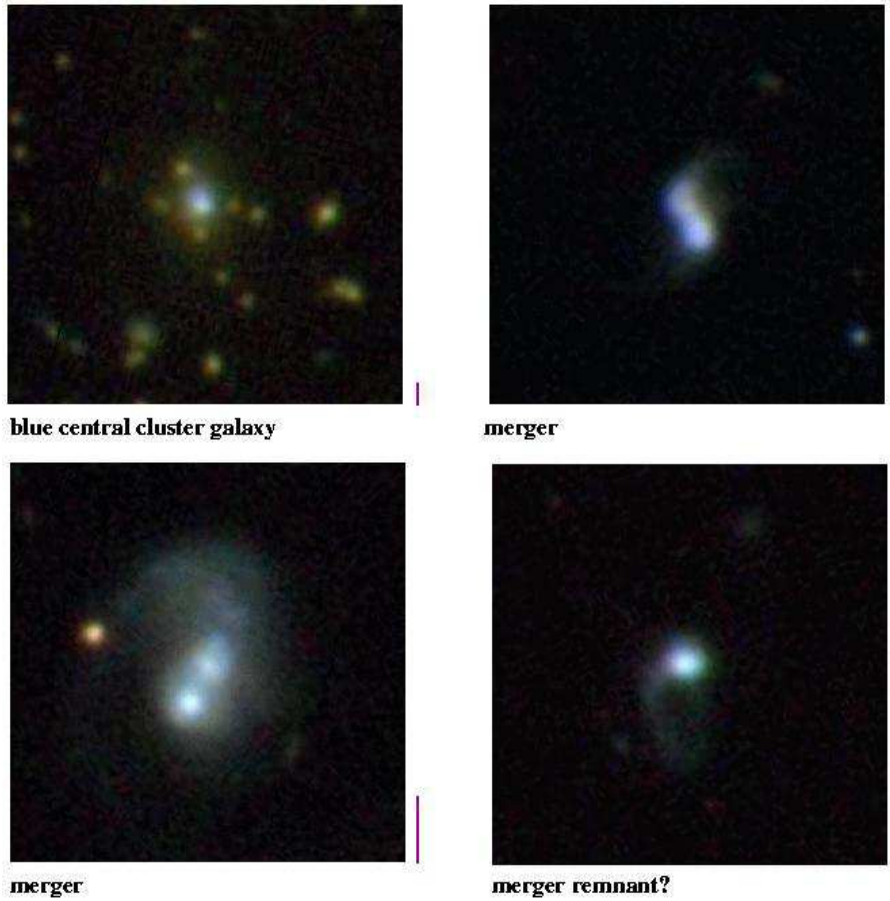}
}
\caption{\label{fig25}
\small
A montage of 48"$\times$48"  images of AGN with [OIII] luminosities 
greater than $10^{7.8} L_{\odot}$.}
\end {figure}
\normalsize

The images in Fig. 24 and 25 suggest that many different physical processes may be responsible
for fuelling black holes and building bulges. 
If this is the case, it becomes even
more of a challenge  to understand why there is such a tight connection between
black hole mass and bulge mass.

The presence in our sample  of a substantial number of spheroidal galaxies with apparently
normal structure but young stellar populations
suggests that powerful
Type 2 AGN, like quasars, may be a late stage of an merger/starburst.                      
In section 4.5 we found that many powerful type 2 AGN have radial gradients
in their stellar populations, with the stars near the nucleus exhibiting    
larger 4000 \AA\ break strengths than the stars a few kiloparsecs away.
This can be explained if the {\em mass fraction}
of stars formed in the burst increases as a function of radius. 
It would then take substantially longer for the 4000 \AA\ break to
revert back to values characteristic of normal ellipticals in the main bodies of galaxies
than in their nuclei.
Begelman (1985) has proposed that x-rays from the active galactic nucleus
can produce runaway heating of the ISM and result in central ``holes'' in the distribution
of cold gas in active galaxies. The AGN itself could  thus regulate
the amount of star formation that occurs at the centres of galaxies.

Quite apart from any specific scenario, our principal result 
has an appealingly simple interpretation. The two
necessary ingredients for a powerful AGN are a massive black hole
and an abundant fuel supply.
Only massive early-type galaxies contain massive
black holes, and only galaxies with significant amounts of recent/on-going
star-formation have the requisite fuel supply. This combination is rare
today, but evidently was not so at high redshift.

Further clues to the connection between type 2 AGN, QSOs, ULIRGs and radio galaxies 
 will come from direct
cross-correlation of our sample with infrared and radio  surveys such as IRAS and FIRST
(e.g. Ivezi\'{c} et al. 2002). In addition, the SDSS offers
unprecedented opportunities for a detailed comparison of the environments of AGN with those
of normal galaxies. Our analysis has demonstrated a clear link between 
type 2 AGN and the formation or evolution of
massive early-type galaxies. The challenge is now to understand  this connection in more detail.\\
   
\vspace {1.0cm}
S.C. thanks the Alexander von Humboldt Foundation, the Federal 
Ministry of Education and Research, and the Programme for Investment
in the Future (ZIP) of the German Government for their support.

Funding for the creation and distribution of the SDSS Archive has been provided
by the Alfred P. Sloan Foundation, the Participating Institutions, the National
Aeronautics and Space Administration, the National Science Foundation, the U.S.
Department of Energy, the Japanese Monbukagakusho, and the Max Planck Society.
The SDSS Web site is http://www.sdss.org/.

The SDSS is managed by the Astrophysical Research Consortium (ARC) for the
Participating Institutions. The Participating Institutions are The University
of Chicago, Fermilab, the Institute for Advanced Study, the Japan Participation
Group, The Johns Hopkins University, Los Alamos National Laboratory, the
Max-Planck-Institute for Astronomy (MPIA), the Max-Planck-Institute for
Astrophysics (MPA), New Mexico State University, University of Pittsburgh,
Princeton University, the United States Naval Observatory, and the University
of Washington.\\

\vspace{1.5cm}

\Large
\begin {center} {\bf References} \\
\end {center}
\normalsize
\parindent -7mm
\parskip 3mm

Adams, T.F., 1977, ApJS, 33, 19

Antonucci, R. 1993, ARA\&A,   31, 473

Bahcall, J.N., Kirhakos, S., Saxe, D.H., Schneider, D.P., 1997, ApJ, 479, 624

Balogh, M.L., Morris, S., Yee, H., Carlberg, R., \& Ellingson, E.
1999, ApJ, 527, 54

Baldwin, J., Phillips, M., \& Terlevich, R. 1981, PASP, 93, 5

Begelman, M., 1985, ApJ, 297, 492

Blanton, M.R., Lupton, R.H., Maley, F.M., Young, N., Zehavi, I., 
Loveday, J. 2003, AJ, in press (astro-ph/0105535)

Bruzual, G., \& Charlot, S. 2003, MNRAS, submitted

Canalizo, G., Stockton, A., 2001, ApJ, 555, 719

Charlot, S., Fall, 2000, ApJ, 539, 718 

Charlot, S., Longhetti, M., 2001, MNRAS,323,887

Cid Fernandes, R., Heckman, T., Schmitt, H., Gonz\'alez Delgado, R.,
Storchi-Bergmann, T. 2001, ApJ, 558, 81

Croom, S.M., et al, 2002, MNRAS, 335, 459

Dahari, O., De Robertis, M.M., 1988, ApJ, 331, 727

De Grijp, M.H.K., Keel, W.C., Miley, G.K., Goudfrooij, P., Lub, J., 1992, A\&AS, 96, 389

De Robertis, M.M., Hayhoe, K., Yee, H.K.C., 1998, ApJS, 115, 163

Dunlop, J., McLure, R., Kukula, M., Baum, S., O'Dea, C., \& Hughes,
D. 2001, astro-ph/0108397

Ferrarese, L.,  Merritt, D. 2000, ApJ,  539, L9

Fukugita, M., Ichikawa, T., Gunn, J.E., Doi, M., Shimasaku, K., 
Schneider, D.P. 1996, AJ, 111, 1748

Gebhardt, K., Bender, R., Bower, G., Dressler, A., Faber, S.,
Filippenko, A., Green, R., Grillmair, C., Ho, L., Kormendy, J.,
Lauer, T., Magorrian, J., Pinkney, J., Richstone, D., Tremaine, S. 2000,
ApJ, 539, L13

Genzel, R., Tacconi, L.J., Rigopoulou, D., Lutz, D., Tecza, M., 2001, ApJ, 563, 527

Gonz\'alez Delgado, R., Heckman, T., Leitherer, C. 2001, ApJ, 546, 845

Gunn, J., Carr, M., Rockosi, C., Sekiguchi, M., Berry, K., Elms,
B., de Haas, E., Ivezi\'{c}, Z. et al. 1998, ApJ, 116, 3040

Haehnelt, M.G.,  Natarajan, P., Rees, M.J., 1998, MNRAS, 300, 817

Heckman, T.M., 1978, PASP, 90, 241

Heckman, T.M. 1980a, A\&A , 87, 142

Heckman, T.M. 1980b, A\&A , 87, 152

Heckman, T.M., 2003, Proceedings of the Carnegie Symposium on the Co-Evolution
of Black Holes and Galaxies, ed Ho, L.

Ho, L.C., Filippenko, A.V., Sargent, W.L.W., 1993, ApJ, 417, 63

Ho, L.C. Filippenko, A.V., Sargent, W.L.W., 1995, ApJS, 98, 477

Ho, L.C., Filippenko, A.V., Sargent, W.L.W., 1997, ApJS, 112, 315

Ho, L.C., Filippenko, A.V., Sargent, W.L.W., 2003, ApJ, 583, 159 

Hogg, D., Finkbeiner, D., Schlegel, D., \& Gunn, J. 2001, AJ, 122, 2129

Huchra, J., Burg, R., 1992, ApJ, 393, 90

Ivezi\'{c}, Z., Menou, K., Knapp, G.R., Strauss, M.A., Lupton, R.H., Vandenberk, D.E.,
Richards, G.T., Tremonti, C. et al, 2002, AJ, 124, 2364

Joguet, B., Kunth, D., Melnick, J., Terlevich, R., \& Terlevich, E. 2001,
A\&A, 380, 19

Kauffmann, G.,  Haehnelt, M.G.,  2000, MNRAS,  311, 576
                           
Kauffmann, G., Heckman, T., White, S., Charlot, S., Tremonti, C.,
Brinchmann, J., Bruzual, G., Peng, E., Seibert, M. et al. 2003a, MNRAS, 
in press (astro-ph/0204055, Paper I)
                           
Kauffmann, G., Heckman, T., White, S., Charlot, S., Tremonti, C.,
Peng, E., Seibert, M., Brinkmann, J., et al.
2003b, MNRAS, in press (astro-ph/0205070, Paper II)

Kewley, L., Dopita, M., Sutherland, R.,
Heisler, C., Trevena, J. 2001, ApJ, 556, 121

Koski, A. 1978, ApJ, 223, 56

Kotilainen, J.K., Ward, M.J., 1994, MNRAS, 266, 953

Kraemer, S.B., Ho, L.C., Crenshaw, D., Shields, J.C., Filippenko, A.V., 1999, ApJ, 520,564

Ledlow, M.J., Owen, F.N., 1996, AJ, 112, 9

Lynden-Bell, D., 1969, Nature, 223, 690

Maiolino, R., Ruiz, M., Rieke, G., \& Keller, L. 1995, ApJ, 446, 561

Malkan, M., Gorjian, V.,  Tam, R. 1998, ApJS, 117, 25

McLure, R.J., Kukula, M.J., Dunlop, J.S., Baum, S.A., O'Dea, C.P., Hughes, D.H., 1999, MNRAS, 308, 377

McLure, R.J., Dunlop, J.S., Kukula, M.J., 2000, MNRAS, 318, 693

Monaco, P., Salucci, P., Danese, L., 2000, MNRAS, 311, 279

Nelson, C.H., Whittle, M., 1996, ApJ, 465, 96

Nolan, L.A., Dunlop, J.S., Kukula, M.J., Hughes, D.H., Boroson, T., Jiminez, R., 2001, MNRAS, 323, 308

Oliva, E., Origlia, L. Maiolino, R., \& Moorwood, A. 1999, A\&A , 350, 9

Osterbrock, D.E., 1989, ``Astrophysics of gaseous nebulae and active galactic nuclei'' 

Percival, W.J., Miller, L., McLure, R.J., Dunlop, J.S., 2001, MNRAS, 322, 843

Pier, J.R., Munn, J.A., Hindsley, R.B., Hennessy, G.S., Kent, S.M.,
Lupton, R.H., Ivezi\'{c}, Z., 2003, AJ, 125, 1559

Richards, G., Fan, X.,  Newberg, H., Strauss, M., Vanden Berk, D.,
Schneider, D., Yanny, B. et al. 2002, AJ, 123, 2945

Richstone, D., Ajhar, E.A., Bender, R., Bower, G., Dressler, A., Faber, S.M., Filippenko, A.V.,
Gebhardt, K., Green, R., Ho, L.C., et al., 1998, Nature, 395, 14

Sanders, D.B., Soifer, B.T., Elias, J.H., Madore, B.F., Matthews, K., Neugebauer, G.,
Scoville, N.Z., 1988, ApJ, 325, 74

Schmitt, H.R., Storchi-Bermann, T., Fernandes, R.C., 1999, MNRAS, 303, 173  

Smith, J.A., et al 2002, AJ, 123, 2121

Stoughton, C., Lupton, R., Bernardi, M., Blanton, M., Burles, S.,
Castander, F., Connolly, A., Eisenstein, D. et al. 2002, AJ, 123, 485

Strateva, I.,  Ivezi\'{c}, Z., Knapp, G., Narayanan, V.,  Strauss, M.,
Gunn, J., Lupton, R., Schlegel, D. et al. 2001, AJ, 122, 1104

Strauss, M. Weinberg, D., Lupton, R., Narayanan, V., Annis, J.,Bernardi,
M., Blanton, M., Burles, S. et al. 2002, AJ, 124, 1810

Tacconi, L.J., Genzel, R., Lutz, D., Rigopoulou, D., Baker, A.J., Iserlohe, C.,
Tecza, M., 2002, ApJ, 580, 73

Ulvestad, J.S., Ho, L.C., 2002, ApJ, 581, 925

Vanden Berk, D.E., Richards, G.Y., Bauer, A., Strauss, M.A., Schneider, D.P., Heckman, T.M.,
York, D.G., Hall, P.B. et al, 2001, AJ, 122, 549

Veilleux, S.,  Osterbrock, D. 1987, ApJS, 63, 295

Worthey, G.,  Ottaviani, D.L. 1997, ApJS, 111, 377

York, D.G. et al. 2000, AJ, 120, 1579

\end{document}